\documentclass[twocolumn]{aastex63}
\usepackage{amsmath}
\usepackage{comment}
\usepackage{graphicx}
\usepackage{xcolor}
\usepackage{xparse,xcoffins}
\received{June 1, 2019}
\revised{January 10, 2019}
\accepted{\today}
\submitjournal{APJ}

\shorttitle{LIM mock}
\shortauthors{Yang et al.}
\usepackage{tgtermes}
\newcommand{\hi}{\ion{H}{1}}
\newcommand{\hii}{\ion{H}{2}}
\newcommand{\ci}{\ion{C}{1}}
\newcommand{\cii}{\ion{C}{2}}

\ExplSyntaxOn
\NewCoffin\imagecoffin
\NewCoffin\labelcoffin

\keys_define:nn { miguel/label }
 {
  label   .tl_set:N = \l_miguel_label_tl,
  labelbox .bool_set:N = \l_miguel_label_box_bool,
  labelbox .default:n = true,
  fontsize .tl_set:N = \l_miguel_label_size_tl,
  fontsize .initial:n = \footnotesize,
  pos .choice:,
  pos/nw .code:n = \tl_set:Nn \l_miguel_label_pos_tl { left,up },
  pos/ne .code:n = \tl_set:Nn \l_miguel_label_pos_tl { right,up },
  pos/sw .code:n = \tl_set:Nn \l_miguel_label_pos_tl { left,down },
  pos/se .code:n = \tl_set:Nn \l_miguel_label_pos_tl { right,down },
  pos/n .code:n = \tl_set:Nn \l_miguel_label_pos_tl { hc,up },
  pos/w .code:n = \tl_set:Nn \l_miguel_label_pos_tl { left,vc },
  pos/s .code:n = \tl_set:Nn \l_miguel_label_pos_tl { hc,down },
  pos/e .code:n = \tl_set:Nn \l_miguel_label_pos_tl { right,vc },
  pos .initial:n = nw,
  unknown .code:n   = \clist_put_right:Nx \l_miguel_label_clist
                       { \l_keys_key_tl = \exp_not:n { #1 } }
 }
\clist_new:N \l_miguel_label_clist
\box_new:N \l_miguel_label_box
\box_new:N \l_miguel_label_image_box

\NewDocumentCommand{\xincludegraphics}{O{}m}
 {
  \group_begin:
  \tl_clear:N \l_miguel_label_tl
  \clist_clear:N \l_miguel_label_clist
  \keys_set:nn { miguel/label } { #1 }
  \tl_if_empty:NTF \l_miguel_label_tl
   {
    \miguel_includegraphics:Vn \l_miguel_label_clist { #2 }
   }
   {
    \SetHorizontalCoffin\imagecoffin
     {
      \miguel_includegraphics:Vn \l_miguel_label_clist { #2 }
     }
    \SetHorizontalCoffin\labelcoffin
     {
      \raisebox{\depth}
       {
        \bool_if:NTF \l_miguel_label_box_bool
         { \fcolorbox{white}{white}{\l_miguel_label_size_tl\l_miguel_label_tl} }
         { \l_miguel_label_size_tl\l_miguel_label_tl }
       }
     }
    \SetVerticalPole\imagecoffin{left}{3pt+\CoffinWidth\labelcoffin/2}
    \SetVerticalPole\imagecoffin{right}{\Width-3pt-\CoffinWidth\labelcoffin/2}
    \SetHorizontalPole\imagecoffin{up}{\Height-3pt-\CoffinHeight\labelcoffin/2}
    \SetHorizontalPole\imagecoffin{down}{3pt+\CoffinHeight\labelcoffin/2}
    \use:x{\JoinCoffins\imagecoffin[\l_miguel_label_pos_tl]\labelcoffin[vc,hc]} 
    \TypesetCoffin\imagecoffin
   }
   \group_end:
 }
\NewDocumentCommand{\setlabel}{m}
 {
  \keys_set:nn { miguel/label } { #1 }
 }

\cs_new_protected:Nn \miguel_includegraphics:nn
 {
  \includegraphics[#1]{#2}
 }
\cs_generate_variant:Nn \miguel_includegraphics:nn { V }

\ExplSyntaxOff

\begin{document}

\title{multi-tracer cosmological line intensity mapping mock lightcone simulation}

\correspondingauthor{Shengqi Yang}
\email{sy1823@nyu.edu}

\author{Shengqi Yang}
\affiliation{Center for Cosmology and Particle Physics, Department of physics, New York University, 726 Broadway, New York, NY, 10003, U.S.A.}

\author{Rachel S. Somerville}
\affiliation{Center for Computational Astrophysics, Flatiron institute, New York, NY 10010, U.S.A.}
\affiliation{Department of Physics and Astronomy, Rutgers University, 136 Frelinghuysen Road,   Piscataway, NJ 08854}

\author{Anthony R. Pullen}
\affiliation{Center for Cosmology and Particle Physics, Department of physics, New York University, 726 Broadway, New York, NY, 10003, U.S.A.}
\affiliation{Center for Computational Astrophysics, Flatiron institute, New York, NY 10010, U.S.A.}
\author{Gerg\"o Popping}
\affiliation{European Southern Observatory, Karl-Schwarzschild-Strasse 2, D-85748, Garching, Germany}
\author{Patrick C. Breysse}
\affiliation{Center for Cosmology and Particle Physics, Department of physics, New York University, 726 Broadway, New York, NY, 10003, U.S.A.}
\affiliation{Center for Computational Astrophysics, Flatiron institute, New York, NY 10010, U.S.A.}
\author{Abhishek S. Maniyar}
\affiliation{Center for Cosmology and Particle Physics, Department of physics, New York University, 726 Broadway, New York, NY, 10003, U.S.A.}



\begin{abstract}

 Sub-millimeter emission lines are important tracers of the cold gas and ionized environments of galaxies and are the targets for future line intensity mapping surveys. Physics-based simulations that predict multiple emission lines arising from different phases of the interstellar medium are crucial for constraining the global physical conditions of galaxies with upcoming line intensity mapping observations. In this work we present a general framework for creating multi-tracer mock sub-millimeter line intensity maps based on physically grounded galaxy formation and sub-mm line emission models. We simulate a mock lightcone of 2 deg$^2$ over a redshift range $0\leq z\leq10$, comprising discrete galaxies and galaxy [\cii], CO, [\ci] emission. We present simulated line intensity maps for two fiducial surveys with resolution and observational frequency windows representative of COMAP and EXCLAIM. We show that the star formation rate and line emission scaling relations predicted by our simulation significantly differ at low halo masses from widely used empirical relations, which are often calibrated to observations of luminous galaxies at lower redshifts. We show that these differences lead to significant changes in key summary statistics used in intensity mapping, such as the one point intensity probability density function and the power spectrum. It will be critical to use more realistic and complex models to forecast the ability of future line intensity mapping surveys to measure observables such as the cosmic star formation rate density.

\end{abstract}



\section{Introduction} \label{sec:intro}

Cosmic microwave background experiments and galaxy surveys have made great progress in advancing our understanding of the origin and evolution of different components of our Universe. However, the wealth of data provided by these measurements raises more profound questions. For example, whether the observed accelerating expansion of the Universe should be explained by the presence of dark energy or the breakdown of general relativity on cosmological scale is still under debate \citep{1999ApJ...517..565P,2000PhLB..485..208D,2003RvMP...75..559P,2004PhRvD..70d3528C}. On the astrophysics side, the cause of the cosmic galaxy star formation rate (SFR) density deviating from the continuous growth of dark matter (DM) halos from redshift $z\sim2$ to the present is still unclear \citep{2014ARA&A..52..415M}. Moreover, our knowledge about the conditions in the interstellar medium (ISM) and the effects on the global properties of galaxies is limited, which is crucial for understanding the star formation (SF) and galaxy evolution process \citep{2013ARA&A..51..105C}. Answering these questions requires measurements with higher resolution, larger observational volumes or new experimental designs.\par 
Line intensity mapping (LIM) is an emerging technique to advance our understanding about both cosmology and extragalactic astronomy in the next decade \citep{2017arXiv170909066K}. Unlike galaxy surveys which resolve individual sources, LIM integrates all the emission along the line of sight, including the signal contributed by faint sources. Advantages of LIM are three-fold. First, LIM can probe vast cosmological volumes at high redshifts, where the emitters are generally too faint to be resolved in galaxy surveys. This feature not only allows tests of gravity on cosmological scales, but also provides information about the large-scale structure distributions from the present time all the way back to the epoch of reionization (EoR). Secondly, since all the emission along the line of sight is collected in the LIM experiments, LIM surveys sample the entire galaxy population, while traditional galaxy surveys are biased towards the brightest sources. This feature is important for inferring ISM and SF properties at different cosmic eras. Lastly, by not attempting to resolve individual sources, LIM only requires a modest telescope aperture size, which is more economical. \par 
LIM was originally conceived to map the 21-cm line, an important tracer of the matter distribution, emitted by neutral hydrogen during the EoR as well as the post-reionization epoch. The molecular and fine structure lines emitted from galaxies, such as [\cii] and CO lines have also attracted interest, leading to the design of several LIM experiments. Since different emission lines are unique tracers of the corresponding gas phases, multi-tracer LIM studies will provide new information about ISM conditions for various cosmic times. In the next decade, numerous LIM surveys will be conducted. Some examples are HIRAX \citep{2016SPIE.9906E..5XN}, CHIME \citep{2017ApJ...844..161A} and HERA \citep{2017PASP..129d5001D} which probe 21 cm; SPHEREx \citep{2014arXiv1412.4872D} targeting Ly$\alpha$ and H$\alpha$ lines; future CO LIM experiments COMAP \citep{2016ApJ...817..169L}; TIME \citep{2014SPIE.9153E..1WC}, TIM \citep{2018JLTP..193..968H}, CONCERTO \citep{2019MNRAS.485.3486D}, and EXCLAIM \citep{2020JLTP..199.1027A} measuring [\cii] emission.\par 
Two unsolved problems for LIM survey data analysis are 1) disentangling the target emission line signal from the Milky Way (MW) foreground, cosmic infrared background (CIB), interloper lines and other contamination and 2) connecting the measured LIM signal to the physical properties of line emitters as well as cosmological quantities of interest. Analytic models and numerical simulations are powerful tools to help answer these questions. Compared to numerical simulations, analytic models of line emission enjoy the advantage of higher computational efficiency, but suffer from many limitations. First, empirical models are often calibrated only to local galaxy measurements; thus, they cannot be extended confidently to higher redshifts. Theoretical line emission models which are derived from the statistical balance equation generally make assumptions about the ISM properties and only consider single line emission; thus, estimating the cross-correlations between multiple lines becomes challenging and requires extra assumptions about the correlation index. Numerical hydrodynamic cosmological simulations are based on a more robust underlying physics model, but due to the computational cost, they still must make trade-offs between volume and resolution. For example, the high-resolution FIRE simulations \citep{2014MNRAS.445..581H} resolve galaxies with stellar masses down to $10^4M_\odot$, and can partially resolve the multi-phase ISM, but each FIRE simulation only represents one DM halo. Large volume hydrodynamical simulations such as IllustrisTNG \citep{2019ComAC...6....2N}, EAGLE \citep{2016A&C....15...72M}, and SIMBA \citep{2019MNRAS.486.2827D} can typically represent volumes of 50-300 Mpc on a side, but the mass resolution is limited to $\sim10^6-10^9 M_\odot$ and phenomenological ``sub-grid'' models must be used to treat processes such as star formation, stellar feedback, and black hole growth and feedback (see \cite{2015ARA&A..53...51S} for a review). The Quick Particle-Mesh (QPM) mocks used in eBOSS \citep{2014MNRAS.437.2594W} are simulated in a huge box with 2.56 Gpc/h per side, while the DM halo mass resolution is 10$^{12}h^{-1}M_\odot$ -- insufficient to resolve individual galaxies. In all cases, a single cosmological simulation cannot provide information about the large-scale structure ($\sim$ 100 Mpc) and the conditions in the ISM that influence line emission ($\sim$ pc) simultaneously.\par 
One option to cut down the computational expense is to apply empirical relations between DM halos and galaxy properties, such as SFR, and then use empirically calibrated scaling relations to translate this to line emission. This approach has been used quite extensively in the literature ( e.g. \cite{2015MNRAS.450.3829Y,2015ApJ...806..209S,2016ApJ...833..153S,2016ApJ...817..169L,2017MNRAS.464.1948F,2019ApJ...871...75I,2019MNRAS.488.3014P}). 
The drawbacks to this method, as mentioned before, are that the empirical models are calibrated to observations over a certain luminosity and redshift range, but are applied over broader ranges in these quantities, over which the models are not well tested. Moreover, empirical line emission models generally focus on a single tracer and do not self-consistently predict multiple lines. It therefore fails to exploit the exciting potential of using multiple tracers for LIM measurements.\par 
An intermediate approach between empirical halo models and numerical simulations is the semi-analytic model (SAM) approach. Similar to an N-Body simulation, a SAM dynamically evolves an ensemble of DM and baryonic components in a cosmological context. To improve the simulation efficiency, the SAM solves the numerically complex, non-linear physical processes involved in DM halo and galaxy evolution through simplified but physically motivated treatments, which are calibrated to more detailed numerical simulations. SAMs adopt simplifying assumptions for evolving DM and baryons, and like large-volume hydrodynamic simulations, generally adopt phenomenological recipes to treat processes such as star formation and stellar feedback. These recipes contain free parameters that are calibrated to match global observational quantities such as the stellar mass function, galaxy gas fractions, mass-metallicity relation, etc. Therefore, although SAMs have been quite successful at matching a broad range of observations over cosmic time, there are remaining uncertainties about whether inaccuracies in model assumptions might be partly compensated by the freedom to tune free parameters.  Studies that compare N-Body/hydro simulations with SAMs show overall agreement in key global quantities and their evolution over cosmic time \citep{2015ARA&A..53...51S}, although there are still significant discrepancies among the gas properties and SF efficiencies \citep{2001MNRAS.320..261B,2012MNRAS.419.3200H,2018MNRAS.474..492M}. \cite{2019ApJ...882..137P} finds tension between observed galaxy H$_2$ masses at high redshift and predictions from both a SAM and IllustrisTNG, but finds fairly close agreement between the theoretical predictions from the two methods. Overall, SAMs represent a promising method for providing mock data for upcoming large surveys and LIM experiments.  
\par 
The SAM we choose in this work is the Santa Cruz SAM developed by \cite{1999MNRAS.310.1087S,2008MNRAS.391..481S,2012MNRAS.423.1992S,2014MNRAS.444..942P,2014MNRAS.442.2398P,2015MNRAS.453.4337S}. The Santa Cruz SAM partitions ISM gas into atomic, molecular and ionized phases, and adopts  an $\mathrm{H}_2$-based SF recipe  motivated by the observed correlation between SFR and molecular gas density \citep{2008AJ....136.2846B,2011ApJ...730L..13B,2011AJ....142...37S}. The Santa Cruz SAM successfully reproduces various key UV/optical galaxy observations for redshift $z<6$ \citep{2012MNRAS.423.1992S,2015MNRAS.453.4337S} and has been shown to be in excellent agreement with available observations from $z\sim 6$--10, as well as the reionization history of the Universe. The \cite{2019MNRAS.483.2983Y,2019MNRAS.490.2855Y,2020arXiv200108751Y,2020MNRAS.tmp..670Y} paper series also utilizes the Santa Cruz SAM to make high redshift predictions for the upcoming James Webb Space Telescope. \par
 
 \cite{2016MNRAS.461...93P,2019MNRAS.482.4906P} has developed a tool that couples to the Santa Cruz SAM (hereafter referred to as the ``sub-mm SAM") to simulate multiple sub-mm line luminosities for each simulated galaxy. This model has been shown to be successful in reproducing available observations of [\cii], CO and [\ci] luminosity versus SFR and stellar mass across various cosmic times back to $z\sim 6$. The combined Santa Cruz SAM and sub-mm SAM pipeline is simultaneously highly computationally efficient, yet grounded in physics, and able to self-consistently predict a broad suite of observable tracers. It is therefore a particularly powerful tool for generating multiwavelength source catalogs, which can be used to generate mock LIM maps.\par

In this work we construct a 2 deg$^2$ lightcone over the redshift range $0\leq z\leq10$ using DM halos from the Small MultiDark-Planck (SMDPL) N-Body simulation \citep{2016MNRAS.457.4340K}. We then use the Santa Cruz SAM to estimate the merger history for each DM halo and simulate the properties and distribution of galaxies within it. We then apply the sub-mm SAM to estimate the [\cii], CO and [\ci] \ line luminosities for each individual galaxy. By integrating along lines of sight along the lightcone, we create synthetic maps over frequency ranges of interest.  To the best of our knowledge, this is the first LIM simulation that self-consistently models multiple far-infrared (FIR) emission lines and links them with UV-optical-NIR properties of a discrete galaxy source catalog. This mock lightcone catalog is directly relevant to various LIM surveys and will be valuable for future LIM survey design and analysis pipeline development. \par

CO emission is an excellent tracer of ISM molecular gas, which is strongly correlated with star formation activity in galaxies \citep{2013ARA&A..51..207B}. Compared with other sub-mm emission lines, the unique advantage of studying CO emission is that the CO molecule simultaneously emits multiple lines from a ``ladder" of rotational transitions. Therefore, cross-correlation between the CO emission lines observed in different frequency channels could help remove uncorrelated foreground and interloper contamination and provide rich information about the CO emitters. Several CO J=1-0 empirical models have been proposed in the last decade \citep{2008A&A...489..489R,2010JCAP...11..016V,2013ApJ...768...15P,2016ApJ...817..169L}. We will compare the Santa Cruz SAM + sub-mm SAM CO J=1-0 predictions with various empirical CO J=1-0 models and study how CO J=1-0 IM statistics vary under different models. In this work we create a mock map with fiducial characteristics similar to the COMAP pathfinder survey, which will probe CO J=1-0 at $z=2.4-2.8$. 

The fine structure line emitted by ionized carbon, [\cii], is another strong tracer of dense gas and SF. [\cii] is the brightest FIR line and contributes 0.1-1\% of the FIR luminosity of the nuclear region of galaxies and has been modeled analytically by many groups (e.g. \cite{2012ApJ...745...49G,2015ApJ...806..209S,2018MNRAS.478.1911P,2019MNRAS.489L..53Y,2019ApJ...887..142S,2020ApJ...892...51C}). We also create a second fiducial mock map with characteristics representative of the EXCLAIM survey \citep{2020JLTP..199.1027A}, which will probe [\cii] emitters at $z=2.5-3.6$.

Two of the most commonly used summary statistics for LIM are the power spectrum and the PDF of intensity values in voxels, also referred to as the one-point intensity PDF or voxel intensity distribution (VID). The VID is a potentially powerful summary statistic based on well-known $P(D)$ analysis methods \citep{1957PCPS...53..764S} for LIM that can provide constraints on the luminosity function of the target line emitters as well as the SFR density \citep{2016MNRAS.457L.127B,2017MNRAS.467.2996B}. \cite{2019PhRvL.123w1105B} further proposed that combining a one point LIM PDF analysis with galaxy surveys, a statistic called the conditional voxel intensity distribution (CVID), can not only constrain physical processes but also remove uncorrelated foregrounds such as the Milky Way continuum emission and the interloper line contamination. \citet{2019ApJ...871...75I} showed that a joint analysis combining both the power spectrum and VID can yield stronger constraints than either approach independently. In this paper we present VID and power spectra predictions computed directly from our fiducial COMAP and EXCLAIM mock maps.

The plan of this paper is as follows: In section \ref{sec:2} we summarize the method used to create the mock lightcone, populate it with sources, and create synthetic maps. Specifically we briefly introduce the SMDPL N-body simulation in Section \ref{sec:2.1}. The method to construct a lightcone from an N-Body simulation is explained in Section \ref{sec:2.2}. We introduce the Santa Cruz SAM in Section \ref{sec:2.3} and the sub-mm SAM in Section \ref{sec:2.4}. The FIR dust emission model is introduced in Section \ref{sec:2.5}. We explain how we make mock intensity maps for the sub-mm lines, FIR emission and Milky Way (MW) continuum emission in section \ref{sec:2.6}. Finally we introduce two fiducial survey designs considered in this work in section \ref{sec:2.7}. In section \ref{sec:3} we summarize the main results and compare them with observations and other empirical models. We conclude in section \ref{sec:4}. Throughout this work, we assume cosmological parameters consistent with the SMDPL simulation: $\Omega_M=0.307$, $\Omega_B=0.048$, $\Omega_\Lambda=0.693$, $\sigma_8=0.829$, $n_s=0.96$, $h=0.678$.

\section{Tools and Methods} \label{sec:2}
\subsection{Small MultiDark-Planck N-body simulation}\label{sec:2.1}
The volume of a 2 deg$^2$ lightcone along the redshift range $0\leq z\leq10$ is about $5.7\times10^7\ \mathrm{(Mpc/h)}^3$. To ensure the statistical independence of each region within the mock lightcone, we require the N-Body simulation volume to be no less than the target lightcone volume. Moreover, we expect that below a critical mass, halos will not be able to accrete or retain significant gas reservoirs, so that only halos with masses larger than this critical mass will be relevant for simulating detectable line emission. We therefore require the halo mass resolution in the N-Body simulation to be at least $10^{10}M_\odot$. A detailed justification of this mass resolution choice is presented in Appendix \ref{appendix:Mmin}. Due to these two aspects of consideration, in this work we choose the SMDPL cosmological N-body simulation for the lightcone construction.\par 
SMDPL contains $3840^3$ DM particles within a (400 Mpc/h)$^3$ cube and simulates the evolution of DM particles from redshift $z=19$ to $z=0$. 117 snapshots are taken at different redshifts during the simulation, with denser sampling in the lower part of the redshift range. SMDPL assumes a standard  $\Lambda$CDM cosmology, with cosmological parameters $\Omega_M=0.307$, $\Omega_B=0.048$, $\Omega_\Lambda=0.693$, $\sigma_8=0.829$, $n_s=0.96$ and $h=0.678$. Among all the Bolshoi/Multidark simulations, SMDPL uses the smallest particle mass $9.6\times10^7\ M_\odot$ and has the highest halo mass resolution of $10^{10}\  M_\odot$. More details can be found in \cite{2016MNRAS.457.4340K}.

\subsection{The dark matter halo lightcone}\label{sec:2.2}
We adopt the method proposed in \cite{2005MNRAS.360..159B} to construct a mock lightcone of DM halos from the SMDPL N-Body simulation. As a brief summary, we first apply periodic boundary conditions to the N-Body data cube. We then randomly select an origin point and a direction for the line-of-sight to cut out a lightcone with a solid angle 2 deg$^2$ and redshift range $0\leq z\leq10$. The redshift of each halo is determined by its comoving distance from the origin and its peculiar velocity along the line-of-sight. Note that the N-Body simulation provides snapshots at a set of redshifts $[z_1,z_2,...z_i,...z_N]$, where $N$ is the number of snapshots. For a halo with redshift $[z_{i-1}+z_i]/2\leq z_{\mathrm{halo}}<[z_i+z_{i+1}]/2$, we read the halo properties from N-Body snapshot $z_i$. 108 out of the 117 SMDPL snapshots in redshift range $0\leq z\leq10$ are used for the dark matter halo lightcone construction. \par
Due to the periodic boundary conditions, DM halos across the mock lightcone will generate a repeating pattern and gain extra spatial correlations. To suppress this replication effect, we randomly shift, rotate and invert the DM halo position in each N-Body 3D catalog copy while stacking them together. Although this random shuffling breaks the continuity of the DM overdensity field and add negative bias to the spatial correlation function, the bias can be accurately estimated for scales smaller than 20\% of the box size \citep{2005MNRAS.360..159B}.
\par 
We adapted the code for the lightcone construction from the one provided in Peter Behroozi's \textsc{universemachine} package \citep{2019MNRAS.488.3143B}\footnote{\url{https://bitbucket.org/pbehroozi/universemachine.git}}.

\subsection{Santa Cruz semi-analytic model}\label{sec:2.3}
In this work, we use the Santa Cruz SAM described in \cite{1999MNRAS.310.1087S,2008MNRAS.391..481S,2012MNRAS.423.1992S,2014MNRAS.444..942P,2014MNRAS.442.2398P,2015MNRAS.453.4337S} to simulate physical and observable properties of galaxies and how they evolve self-consistently over cosmic time. The Santa Cruz SAM is a comprehensive galaxy formation model which uses a simplified but physical treatment of the key processes that shape galaxy evolution. It divides cold gas into ionized, atomic and molecular phases and applies a H$_2$-based star formation (SF) recipe. It also contains a model accounting for DM halo merging history, evolution of sub-halos and galaxy mergers, shock heating and radiative cooling of hot gas within virialized DM halos, supernova feedback, black hole growth and active galactic nucleus (AGN) feedback, photoionization squelching and other processes involved in galaxy evolution. 
The evolution of galaxies is tightly related to the merger history of the DM halo. The halo merger history is commonly represented by ``merger trees", which can either be extracted directly from N-Body simulations or estimated through other semi-analytic formalism. In this work, the Santa Cruz SAM estimates the merger history through a multi-branch tree algorithm based on the extended Press-Schechter (EPS) formalism \citep{1999MNRAS.305....1S}. The advantage of using the EPS formalism over extracting the DM halo merger history from N-Body simulations is that merger trees provided by N-Body simulations have limited mass resolution, while the EPS formalism can extend merger trees to progenitors with arbitrarily small mass. In this work, we record "root halos" down to $M_{\mathrm{root,min}}=10^{10}M_\odot$, and follow merger histories down to 100th of the root mass, or to $10^{10}M_\odot$, whichever is smaller. 
It has been shown in \citet{2014MNRAS.444..942P} that the predictions of the SAM when using N-body based merger trees from the Bolshoi simulation are very similar to the EPS based predictions. 

Before the reionization epoch, the SAM assumes that each DM halo contains hot gas with mass equal to the baryonic mass. After the universe is fully ionized by $z=11$, a fraction of the baryonic mass is allowed to accrete into the halo, based on the filtering mass obtained from hydrodynamic simulations by \citet{2008MNRAS.390..920O}.  
Hot gas then experiences radiative cooling and collapses into the central galaxy, where it is assumed to form a rotationally supported disc. 
The disc size is estimated following \cite{2008ApJ...672..776S} and the radial distribution of the cold gas disc is described by an exponential profile.\par 
In the most up-to-date version of the Santa Cruz SAM \citep{2014MNRAS.442.2398P,2015MNRAS.453.4337S}, cold gas in the galaxy disc is divided into ionized (\hii), atomic (\hi) and molecular (H$_2$) phases. The SFR surface density is modeled by molecular hydrogen-based recipes which are calibrated to observations. \citet{2014MNRAS.442.2398P} and \citet{2015MNRAS.453.4337S} explore a variety of different recipes for gas partitioning and star formation efficiency. They found that the metallicity based, UV-background-dependent recipe based on \cite{2011ApJ...728...88G} (GK model) combined with an H$_2$-based SF relation with a density-dependent slope (Bigiel2 model) gives results that best match observations from the local Universe to $z=4$. \citet{2019MNRAS.483.2983Y,2019MNRAS.490.2855Y,2020arXiv200108751Y,2020MNRAS.tmp..670Y} then confirmed that this model produces the best agreement with observations up to $z\sim 10$. We therefore adopt the GK + Bigiel2 model in this work.\par 
Other ingredients such as stellar feedback, heavy element generation and black hole growth are also included. We refer readers to \cite{2008MNRAS.391..481S,2014MNRAS.442.2398P,2015MNRAS.453.4337S} for more details.  Other galaxy evolution parameters are identical to the values presented in \cite{2015MNRAS.453.4337S,2019MNRAS.482.4906P}.

\par

\subsection{Sub-millimetre emission line modeling}\label{sec:2.4}
The major source of the emission lines we consider in this LIM simulation, i.e. the [\cii], CO and [\ci] \ lines, is dense molecular clouds (MCs) in the ISM. 
The Santa Cruz SAMs predict the scale length of the cold ISM gas in the disk, and the fraction of gas in a dense molecular phase, but do not provide predictions on the properties of MC. 
In this work we adopt the sub-resolution recipe developed in the sub-mm SAM proposed by \cite{2019MNRAS.482.4906P} (hereafter GP19) to simultaneously model multiple sub-mm lines. Specifically, each simulated galaxy is divided into radial annuli, and for each annulus the mass of ionized, atomic and molecular gas are computed following the GK model. The sub-mm SAM then randomly generates MC with masses in the range 10$^4M_\odot<M_{\mathrm{MC}}<10^7M_\odot$ following a power law mass function:
\begin{equation}
    \dfrac{dN}{dM}\propto M^{-\beta}
\end{equation}
GP19 showed that the specific value of $\beta$ does not influence the line emission predictions much. In this work we assume $\beta=1.8$ based on local observations of cloud distribution functions. The random MC generating process stops when the total mass of H$_2$ in the simulated MCs reaches the mass of H$_2$ in the corresponding galaxy annulus. Sub-mm SAM then divides each MC into multiple zones and uses DESPOTIC \citep{2014MNRAS.437.1662K}, a code which solves the energetics of optically thick interstellar clouds, to compute the line emission spectrum. DESPOTIC treats each MC zone as a spherical shell with uniform physical and chemical properties. Given the compositions and physical conditions of a MC zone as well as the external radiation field, DESPOTIC then solves the heating and cooling processes, chemical processes and the profile of spectral lines. Dominant heating processes are the external radiation fields heating and grain photoelectric heating. The main cooling processes are line cooling and dust thermal radiation. Chemical reactions and the corresponding rate coefficients are provided by a reduced carbon-oxygen chemical network \citep{1999ApJ...524..923N} and a non-equilibrium hydrogen chemical network \citep{2007ApJS..169..239G,2012MNRAS.421..116G}. The external ultraviolet (UV) radiation field $G_{\mathrm{UV}}$ and the ionization rate by cosmic rays (CR) $\xi_{\mathrm{CR}}$ are scaled according to the local SFR surface density $\Sigma_{\mathrm{SFR}}$ predicted by the SAM:
\begin{equation}
    \begin{split}
        G_{\mathrm{UV}}&=G_{\mathrm{UV,MW}}\times\dfrac{\Sigma_{\mathrm{SFR}}}{\Sigma_{SFR,MW}}\\
        \xi_{\mathrm{CR}}&=0.1\xi_{\mathrm{CR,MW}}\times\dfrac{\Sigma_{\mathrm{SFR}}}{\Sigma_{\mathrm{SFR,MW}}}
    \end{split}
\end{equation}
Here the MW SFR surface density $\Sigma_{\mathrm{SFR,MW}}=790\, \mathrm{M_\odot Myr^{-1}kpc^{-2}}$ \citep{2011MNRAS.415.2827B}, UV radiation field $G_{\mathrm{UV,MW}}=9.6\times10^{-4}\,\mathrm{erg\, cm^{-2}s^{-1}}$ \citep{2012ApJ...758..109S}, and cosmic ray ionization rate in the diffuse ISM $\xi_{\mathrm{CR,MW}}=10^{-16}\mathrm{s}^{-1}$ \citep{2017MNRAS.467...50N}. Following \cite{2017MNRAS.467...50N}, a factor of 0.1 is introduced to $\xi_\mathrm{CR,MW}$ to account for the cosmic ray shielding in the interiors of MCs. We refer readers to \cite{2014MNRAS.437.1662K} for more details about DESPOTIC and GP19 for the parameters we use to compute the line luminosities.\par 
In this work we grid each MC into 25 zones, which is sufficient for producing convergent [\cii], CO and [\ci] \ luminosities. The density profile assumed for the MCs is another crucial parameter in the line emission simulation. GP19 assumes all the MCs have a Plummer density profile, which was shown to produce a range of line luminosity versus SFR relations in the best agreement with observations available at the time of publication (2018). However, we find the [\cii] luminosities predicted by the original version of sub-mm SAM are lower than more recent ALMA observations by a factor of $\sim3$ at high redshifts. \cite{2020ApJ...890...24V} also suggests that at high redshifts the [\ci] luminosity versus infrared luminosity relations $L_\mathrm{CI}-L_\mathrm{IR}$ predicted by GP19 are significantly lower than observations. Motivated by those discrepancies, we replace the Plummer density profile of molecular clouds by the power-law density profile introdued in GP19. Variation of the cloud radial density profile does not significantly influence the CO luminosity predictions, but it effectively boosts the [\cii] and [\ci] luminosities so that all [\cii], CO and [\ci] scaling relations predicted by sub-mm SAM are in better agreement with current observations. Sub-mm SAM also accounts for the atomic diffuse ISM emission by modeling this ISM phase as one-zone clouds. The hydrogen density and column density of the diffuse atomic gas are fixed as $n_\mathrm{H}$=10 cm$^{-3}$ and $N_\mathrm{H}=10^{21}$ cm$^{-2}$ respectively. Finally, the luminosity of [\cii], CO and [\ci] \ lines emitted from all MC zones and galaxy annuli are summed over to provide the total line luminosity of each galaxy. GP19 showed that this fiducial model produces [\cii], CO and [\ci]\ luminosity versus SFR relations in good agreement with available observations over a broad range of cosmic time. 

\subsection{Dust continuum emission modeling}\label{sec:2.5}
The cosmic infrared background (CIB) signal is the dominant correlated contamination in LIM experiments. Models that account for absorption and emission by dust in the ISM of galaxies are implemented in the Santa Cruz SAM in a manner similar to that described in \cite{2012MNRAS.423.1992S}. 
The Santa Cruz SAM then assumes that all the absorbed energy is re-radiated in the IR and computes the total IR luminosity L$_{\mathrm{IR}}$ of each galaxy. Based on the hypothesis that the dust spectral energy distribution is well correlated with L$_{\mathrm{IR}}$, we use standard dust SED templates to compute the SED spectrum of each galaxy given L$_{\mathrm{IR}}$. Specifically, we first integrate over all the SED templates to compute their total IR luminosity L$^i_{\mathrm{temp}}$ $(i=1,...,N)$, where $N$ is the number of SED templates. We then compare $\log\mathrm{L}_{\mathrm{IR}}$\footnote{In this work $\log$ denotes a base-10 logarithm.} with $\log\mathrm{L}_{\mathrm{temp}}$ and estimate the dust emission SED of each galaxy through linear interpolation. We next integrate the interpolated dust emission SED for each galaxy to compute the IR luminosity in the fiducial survey observable frequency window, which will be specified in Section \ref{sec:2.7}. 
In this work we use dust SED templates provided by \cite{2001ApJ...556..562C}.\par
\begin{figure}
    \centering
    \includegraphics[width=0.45\textwidth]{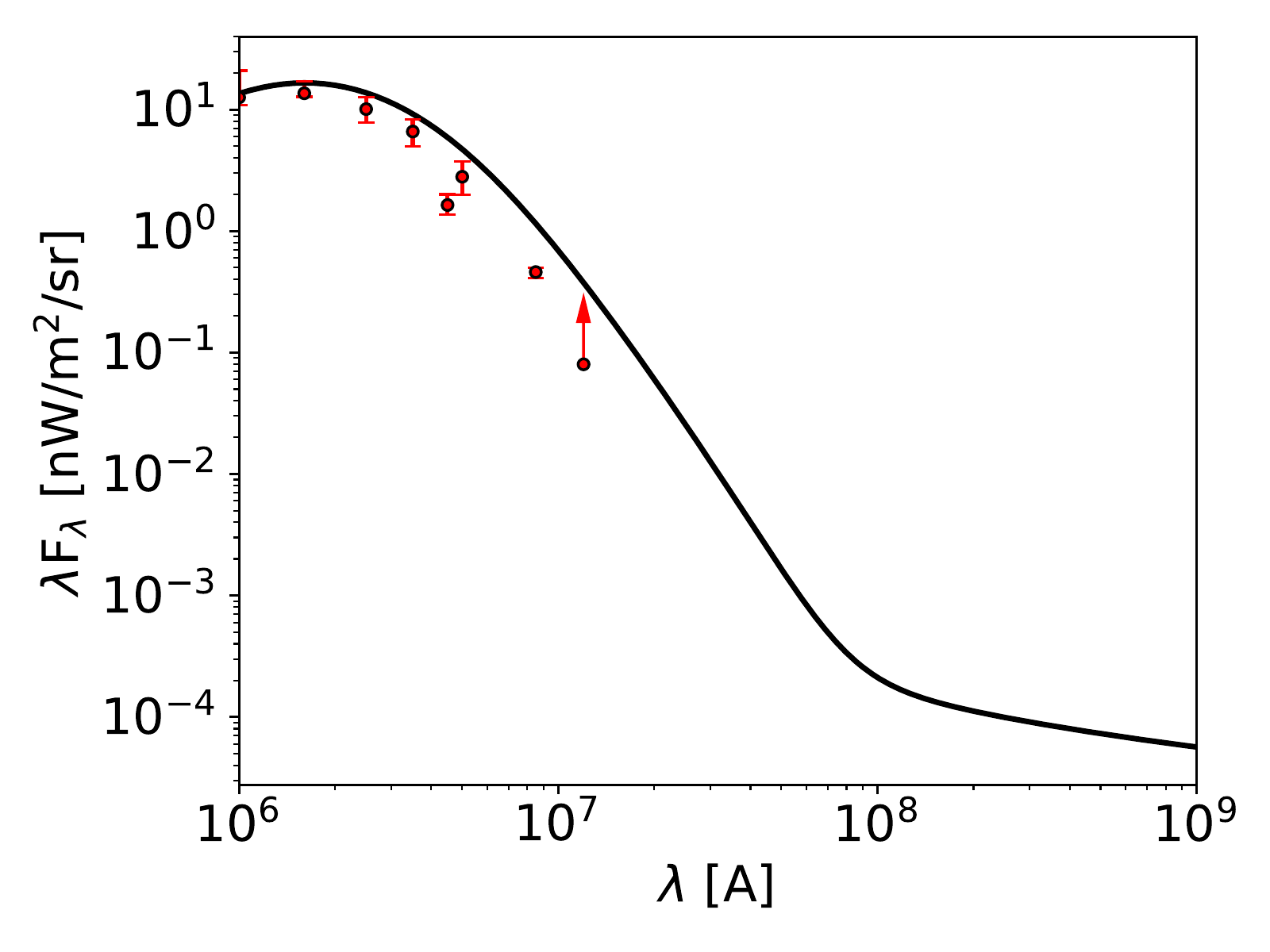}
    \caption{Comparison between the integrated EBL spectrum predicted by the Santa Cruz SAM and observations. The EBL spectrum predicted by the SAM is shown by the black solid curve, while the data points are provided by \cite{2011A&A...532A..49B,2012A&A...542A..58B,2016ApJ...833...71A,2017ApJ...850...37W,2017MNRAS.464.3369Z}, summarized in \cite{2018A&A...614A..39M}. The EBL spectrum predicted by the SAM is consistent with CIB observations over a wide frequency range.}
    \label{fig:1}
\end{figure}

 
We present a comparison between the integrated extragalactic background light (EBL) spectrum of the 2 deg$^2$ mock lightcone predicted by Santa Cruz SAM and observational results in Figure \ref{fig:1}. The observational estimates of the EBL are provided by \cite{2011A&A...532A..49B,2012A&A...542A..58B,2016ApJ...833...71A,2017ApJ...850...37W,2017MNRAS.464.3369Z}, as summarized in \cite{2018A&A...614A..39M}. As shown previously by \cite{2012MNRAS.423.1992S}, this approach produces reasonable agreement with observational EBL constraints. \cite{2012MNRAS.423.1992S} confirmed that halos with mass less than $10^{10}M_\odot$ make negligible contribution to the dust emission.

\subsection{Map making}\label{sec:2.6}
With the sub-mm line modeling and dust continuum emission modeling introduced above, we can construct realistic intensity maps for various target frequencies, including interloper and CIB dust continuum contamination under arbitrary angular and spectral resolution. Below we describe how we create the integrated maps for the emission line and dust continuum emission, include bulk velocities associated with rotation within individual galaxies, and model the finite angular and frequency resolution of an observational map. We also describe how we compute the Milky Way foreground, and the configuration of two fiducial surveys that we will use to construct our maps. 

\subsubsection{Emission line and FIR intensity map}\label{sec:2.6.1}
In this section we provide the details of how we make a mock intensity map. Consider a galaxy at redshift $z$ with disk rotation velocity $v_{\mathrm{disk}}$, emitting one line of interest at rest frame frequency $\nu_0$. Suppose the galaxy inclination angle is $\beta$ (a face-on galaxy corresponds to $\sin\beta=0$, while an edge-on galaxy has $\sin\beta=1$); due to the Doppler effect the emission line profile width increases by
\begin{equation}
    \dfrac{d\nu}{\nu_0}=\dfrac{v_{\mathrm{disk}}\cos\beta}{c}\,,
\end{equation}
where $c$ is the speed of light. We ignore the line broadening caused by thermal motion because the bulk motion is the dominant source of the galaxy velocity dispersion. In this work we use a simple normalized tophat function with width $d\nu/(1+z)$ and mean $\nu_0/(1+z)$ to describe the redshifted line profile $\Phi(\nu)$. The intensity contributed by this single galaxy to a map in the frequency range $[\nu_{\mathrm{min}},\nu_{\mathrm{max}}]$ and angular resolution $\theta_\mathrm{pix}$ is
\begin{equation}\label{eq:5}
    I=\dfrac{L\int_{\nu_{\mathrm{min}}}^{\nu_{\mathrm{max}}}\Phi(\nu)d\nu}{4\pi\chi^2(z)(1+z)^2(\nu_{\mathrm{max}}-\nu_{\mathrm{min}})\Omega_\mathrm{pix}}\,,
\end{equation}
where $L$ is the target line luminosity of the emitter, $\chi(z)(1+z)$ is the luminosity distance, and $\Omega_\mathrm{pix}=\theta^2_\mathrm{pix}$ is the solid angle of each map pixel. The intensity of the FIR emission contributed by a single galaxy at redshift $z$ is computed using a similar method:
\begin{equation}
    I_{\mathrm{CIB}}=\dfrac{\int_{\nu_{\mathrm{min}}}^{\nu_{\mathrm{max}}} F(\nu)d\nu}{4\pi\chi^2(z)(1+z)^2(\nu_{\mathrm{max}}-\nu_{\mathrm{min}})\Omega_\mathrm{pix}}\,,
\end{equation}
here the dust SED $F(\nu)$ is estimated following Section \ref{sec:2.5}.
We repeat this procedure for all the galaxies within the lightcone and sum up all the contributions from different galaxies along the line of sight. 
\par

We grid the 2 deg$^2$ field in the right ascension (RA) and declination (DEC) dimensions with bin width $\theta_\mathrm{pix}$, which is 10 times smaller than the LIM survey beam width $\theta_\mathrm{FWHM}$ (the angular resolution of the fiducial LIM surveys we consider in this work are specified in Section \ref{sec:2.7}, Table \ref{table:1} and Table \ref{table:2}). We treat the spatial distribution of the galaxy light as a delta function for each galaxy, which is a good approximation as our pixels are much larger than the expected extent of individual emitters. We then convolve the maps with a Gaussian with FWHM $\theta_\mathrm{FWHM}$ to represent beam smearing. We additionally grid the 3D intensity maps in the frequency direction with the bin width of the LIM survey frequency resolution.
Pixels near the borders of the Gaussian smeared images are sensitive to the choice of boundary condition. In this work we apply periodic boundary conditions because this choice preserves the LIM total intensity.\par

\subsubsection{Milky Way foreground intensity map}
Cleaning the Milky Way foreground will be one of the great challenges for intensity mapping experiments. Therefore, we provide the option of including a realistic MW foreground in our mock maps. 
We use the Python Sky Model \textsc{PySM} package \citep{2017MNRAS.469.2821T} to simulate the continuum foreground mock map.
The simulated foregrounds are combinations of synchrotron, free-free, anomalous microwave emission and thermal dust emission components. Each component is simulated by the simplest model 1 of \textsc{PySM}. We construct MW foregound maps with $\mathrm{nsides=2048}$ to ensure the angular resolution is higher than the current LIM surveys. Since pixel locations in the full sky maps are different from the other 2 deg$^2$ intensity maps we constructed in Section \ref{sec:2.6.1}, we interpolate the MW foreground intensity on the [RA, DEC] grids using bilinear interpolation and then smooth the interpolated foregrounds to the target angular resolution $\theta_\mathrm{FWHM}$.\par

\subsection{Fiducial surveys}\label{sec:2.7}
In order to provide a demonstration of our map-making tool and compare our simulation with predictions from other models, in this work we make intensity maps for two fiducial surveys. \par
The first fiducial survey is designed to align with COMAP pathfinder \citep{2016ApJ...817..169L} which probes CO J=1-0 emitters at redshift $z=2.4-2.8$. This fiducial survey has observed frequency window 30-34 GHz, spectral resolution $\delta\nu=40$ MHz and angular resolution $\theta_\mathrm{FWHM}=6'$. 

The second fiducial survey is designed to align with EXCLAIM \citep{2020JLTP..199.1027A}, an upcoming balloon mission, which will observe the frequency range 420-540 GHz at spectral resolution $\delta\nu=937.5$ MHz and angular resolution $\theta_{\mathrm{FWHM}}=4'$.\par 
We assume a 2 deg$^2$ sky area centered at RA$=10^\circ$ and DEC$=-35^\circ$ as the mapping region for both fiducial surveys, located within the 408 deg$^2$ EXCLAIM survey area. We model the instrumental noise for COMAP survey as a Gaussian probability density function (PDF) with zero mean and standard deviation $\sigma_\mathrm{N}$. We estimate $\sigma_N$ as the final map sensitivity multiplies $10/\sqrt{8\ln(2)\delta\nu}$ \citep{2016ApJ...817..169L}. Here the $10/\sqrt{8\ln(2)}$ factor is caused by the fact that the map grid in this work is one order smaller than the actual LIM survey beam, and we compute the pixel size as $\theta_\mathrm{pix}^2$ instead of $\theta_\mathrm{pix}^2/(8\ln(2))$, which is slightly different from \citep{2016ApJ...817..169L}. Since the goal of EXCLAIM is to detect the [\cii] statistical anisotropy through cross correlation rather than directly imaging line large scale structure, in this work we only focus on the science signal modeling. We leave a realistic noise model to a dedicated EXCLAIM forecast paper.\par
Parameters of the COMAP and EXCLAIM fiducial surveys are summarized in Table \ref{table:1} and Table \ref{table:2} respectively. The line emission and redshift range of emitters observed in these two fiducial surveys are summarized in Table \ref{table:3}.\par 
\begin{table}
\centering
\begin{tabular}{ |l | c | } 
\hline
Parameter& COMAP fiducial survey  \\ 
\hline
Frequency band ($\Delta\nu$) & 30-34 GHz \\ 
\hline
Frequency channels ($\delta\nu$) & 40 MHz \\ 
\hline
Beam width ($\theta_{\mathrm{FWHM}}$)& 6' \\ 
\hline
Final map sensitivity & 41.5 $\mu$K MHz$^{1/2}$ \\ 
\hline
\end{tabular}
\caption{Summary of parameters for the COMAP pathfinder fiducial survey. Adopted from \cite{2016ApJ...817..169L}.}
\label{table:1}
\end{table}

\begin{table}
\centering
\begin{tabular}{ |l | c | } 
\hline
Parameter& EXCLAIM fiducial survey  \\ 
\hline
Frequency band ($\Delta\nu$) & 420-540 GHz \\ 
\hline
Frequency channels ($\delta\nu$) & 937.5 MHz\\ 
\hline
Beam width ($\theta_{\mathrm{FWHM}}$)& 4' \\ 
\hline
\end{tabular}
\caption{Summary of parameters for the EXCLAIM fiducial survey \citep{2020JLTP..199.1027A}.}
\label{table:2}
\end{table}

\begin{table*}
\centering
\begin{tabular}{ | c | c | c | c | } 
\hline
line& $\nu_0$ [GHz]& redshift range for COMAP fiducial survey & redshift range for EXCLAIM fiducial survey \\ 
\hline
CII & 1901.0  & -& 2.5-3.6\\ 
\hline
CO J=1-0  & 115.3 & 2.4-2.8& -\\ 
\hline
CO J=2-1  & 230.5 & 5.8-6.7& -\\ 
\hline
CO J=3-2  & 345.8 & 9.2-10.0& -\\ 
\hline
CO J=4-3  & 461.0 & -&0.0-0.1 \\ 
\hline
CO J=5-4  & 576.3 & -& 0.0-0.4\\ 
\hline
CI J=1-0  &492.2  & -& 0.0-0.2\\ 
\hline
CI J=2-1  & 809.4 & -& 0.4-1.0\\ 
\hline
\end{tabular}
\caption{Redshift range of line emitters observed in the two fiducial surveys. The second column shows the rest-frame frequencies of emission lines simulated by sub-mm SAM. The third column shows line emitter redshift ranges for the COMAP fiducial survey, for which the observed frequency window is 30-34 GHz. The fourth column shows line emitter redshift ranges for the EXCLAIM 420-540 GHz fiducial survey. ``-" means that the corresponding emission line will not be observable in the frequency window of the relevant fiducial survey.}
\label{table:3}
\end{table*}

\section{Results}\label{sec:3}
\subsection{Geometry and intensity maps}
\begin{figure*}
    \centering
    \xincludegraphics[width=0.9\textwidth]{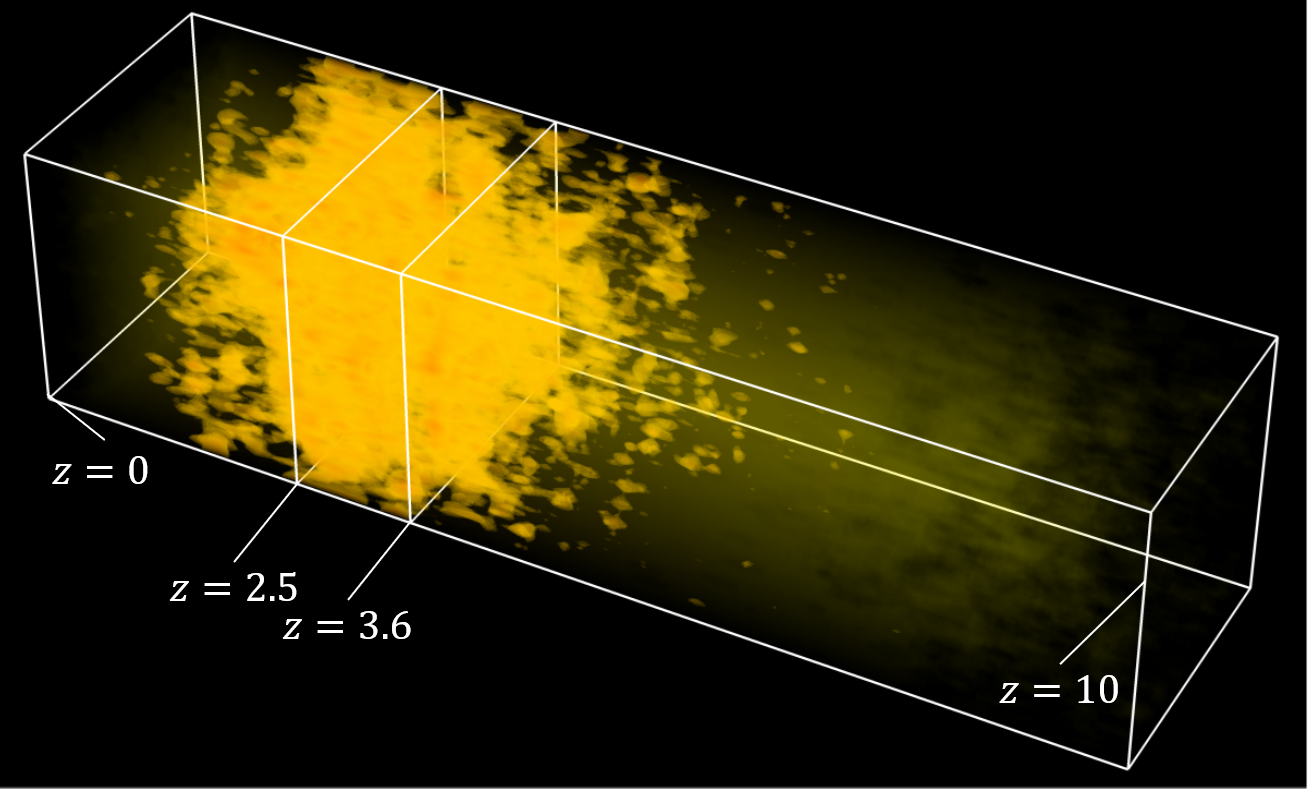}
    \caption{Spatial distribution of DM halos in the mock lightcone. The longest axis shows redshift while the other two axes show RA and DEC of DM halos in the mock. Voxel color is determined by the number counts of DM halos in the corresponding spatial cell. We also highlight the redshift range where the EXCLAIM survey will observe [\cii] emitters   ($2.5<z<3.6$).}
    \label{fig:LCgeometry}
\end{figure*}

The mock lightcone geometry is shown in Figure \ref{fig:LCgeometry}. The longest axis represents redshift while the other two axes show the RA and DEC of each DM halo. The color of each voxel shows the number counts of DM halos with $M_{\mathrm{halo}}>10^{10}M_\odot$ in the corresponding spatial grid. Since the physical volume of the lightcone increases along the redshift direction while the DM halo number density is decreasing, the number counts of DM halos reach a maximum at redshift $2<z<3$ while there are very few halos at $z<1$ and $z>8$. \par
\begin{figure*}
    \includegraphics[width=1\textwidth]{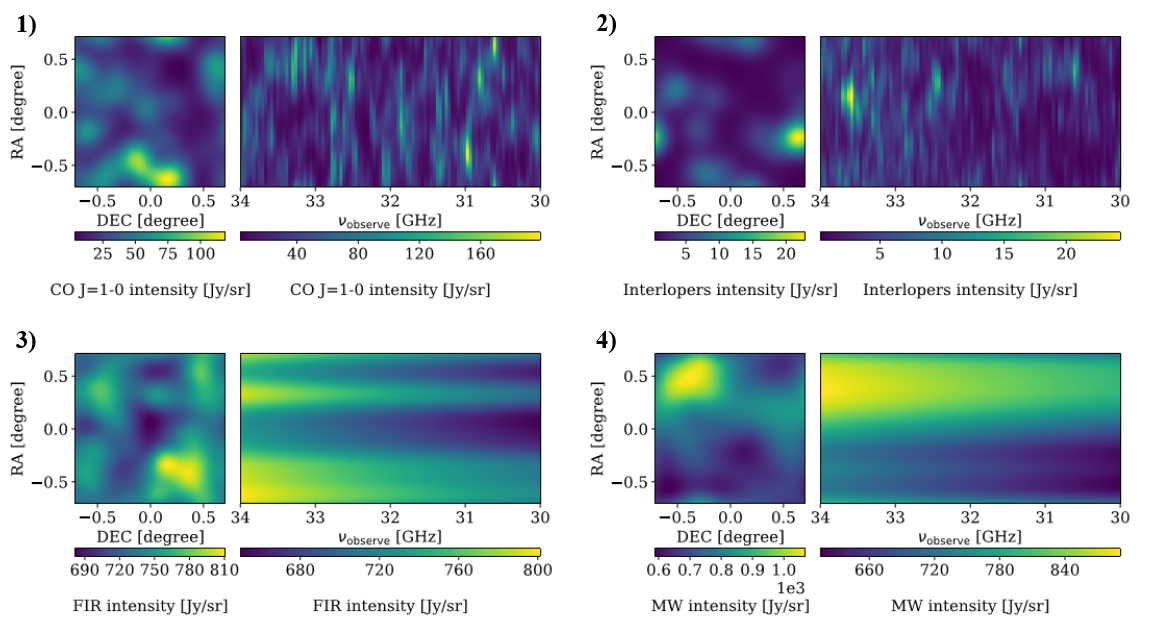}
    \caption{Mock intensity maps for the COMAP pathfinder fiducial survey. The left figure in each panel shows the intensity map at observed frequency 32.00 - 32.04 GHz (``front view"). The right figure shows a ``side view" of the intensity map at DEC= -3' - 3'. 1): CO J=1-0 intensity map . 2) CO J=2-1 and CO J=3-2 interloper line intensity map. 3) CIB intensity map. 4). MW continuum foreground intensity map. 1), 2) and 3) are generated by the Santa Cruz SAM + sub-mm SAM. 4) is generated by the \textsc{PySM} package.}
    \label{fig:IMCOMAP}
\end{figure*}

\begin{figure*}
    \includegraphics[width=1\textwidth]{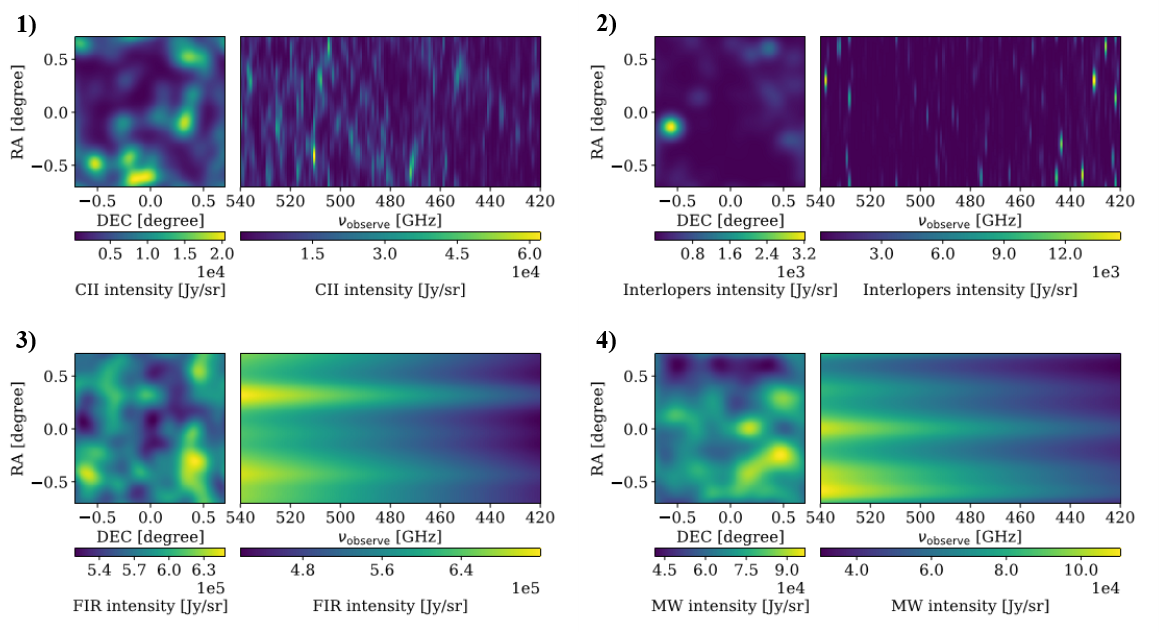}
    \caption{Mock intensity maps for the EXCLAIM fiducial survey. The left figure in each panel shows the intensity map at observed frequency 480.00 - 480.94 GHz (``front view"). The right figure shows a ``side view" of the intensity map at DEC= -2' - 2'. 1): [\cii] intensity map . 2) CO and CI interloper lines intensity map. 3) CIB intensity map. 4). MW continuum foreground intensity map. 1), 2) and 3) are generated by the Santa Cruz SAM + sub-mm SAM. 4) is generated by the \textsc{PySM} package.}
    \label{fig:IMEXCLAIM}
\end{figure*}

We present intensity map slices of the CO J=1-0 ([\cii]) signal, interloper lines and dust continuum background simulated by the Santa Cruz SAM $+$ sub-mm SAM, together with the Milky Way foreground given by \textsc{PySM} for the COMAP (EXCLAIM) fiducial survey in Figure \ref{fig:IMCOMAP} (Figure \ref{fig:IMEXCLAIM}). The CO J=1-0 and [\cii] signals trace the underlying DM density distribution.\par

\subsection{LIM statistics}
In this work we only include halos with masses larger than the N-Body resolution $M_{\mathrm{halo}}>10^{10}M_\odot$ in all statistics.\par
\begin{figure*}
    \centering
    \xincludegraphics[clip,label=1),fontsize=\Large,width=0.49\textwidth]{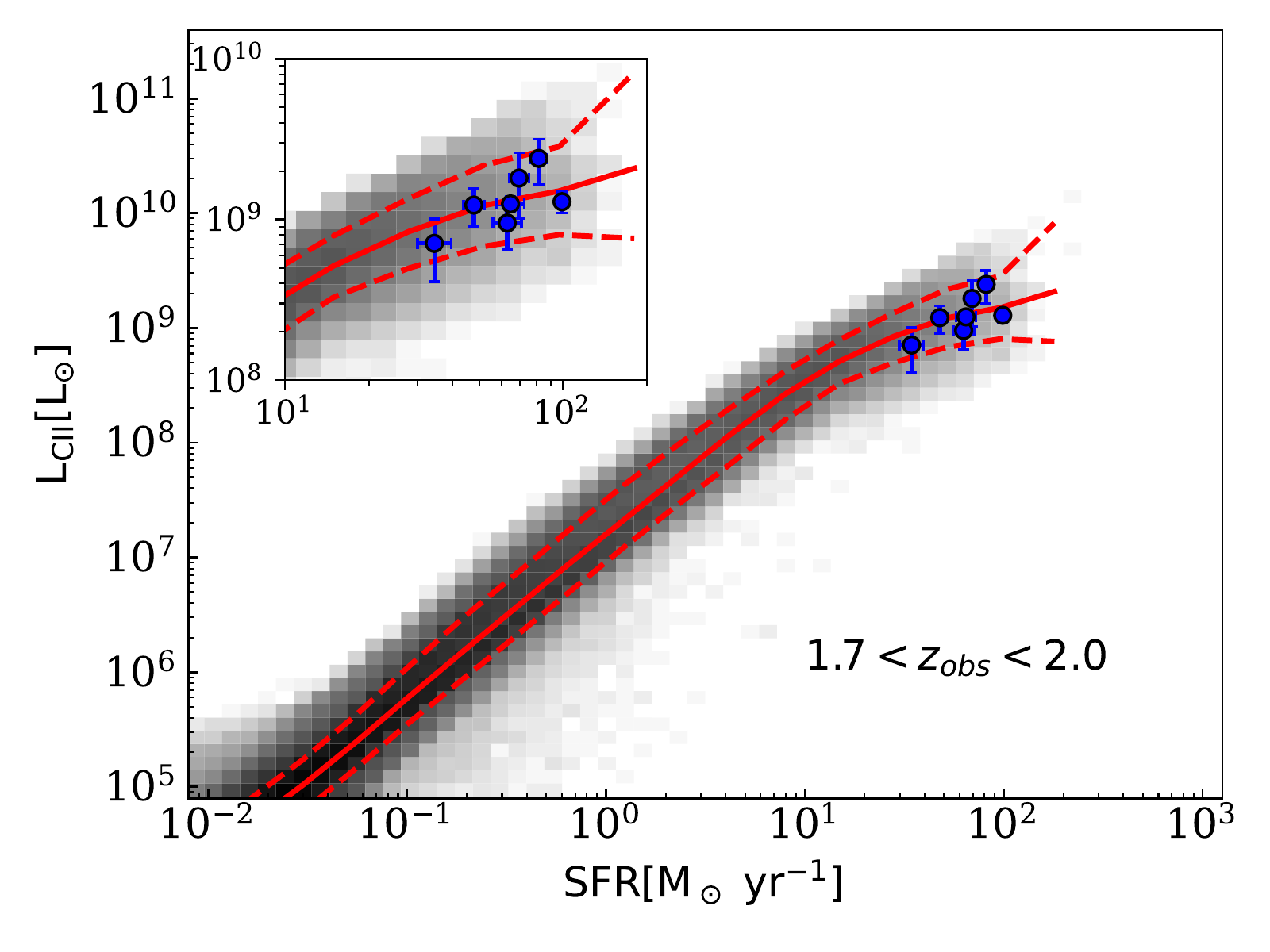}
    \xincludegraphics[clip,label=2),fontsize=\Large,width=0.49\textwidth]{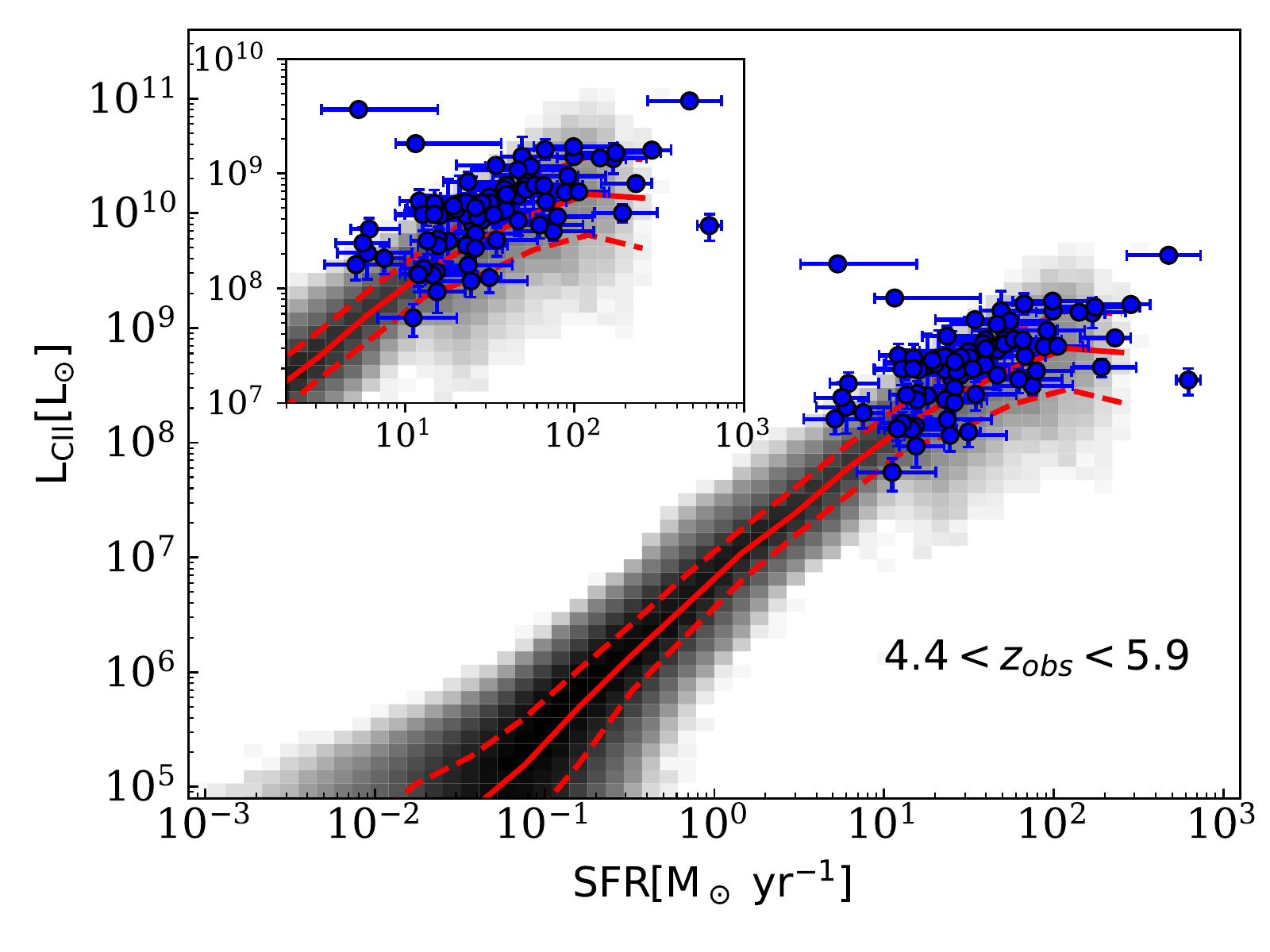}\\
    \xincludegraphics[clip,label=3),fontsize=\Large,width=0.49\textwidth]{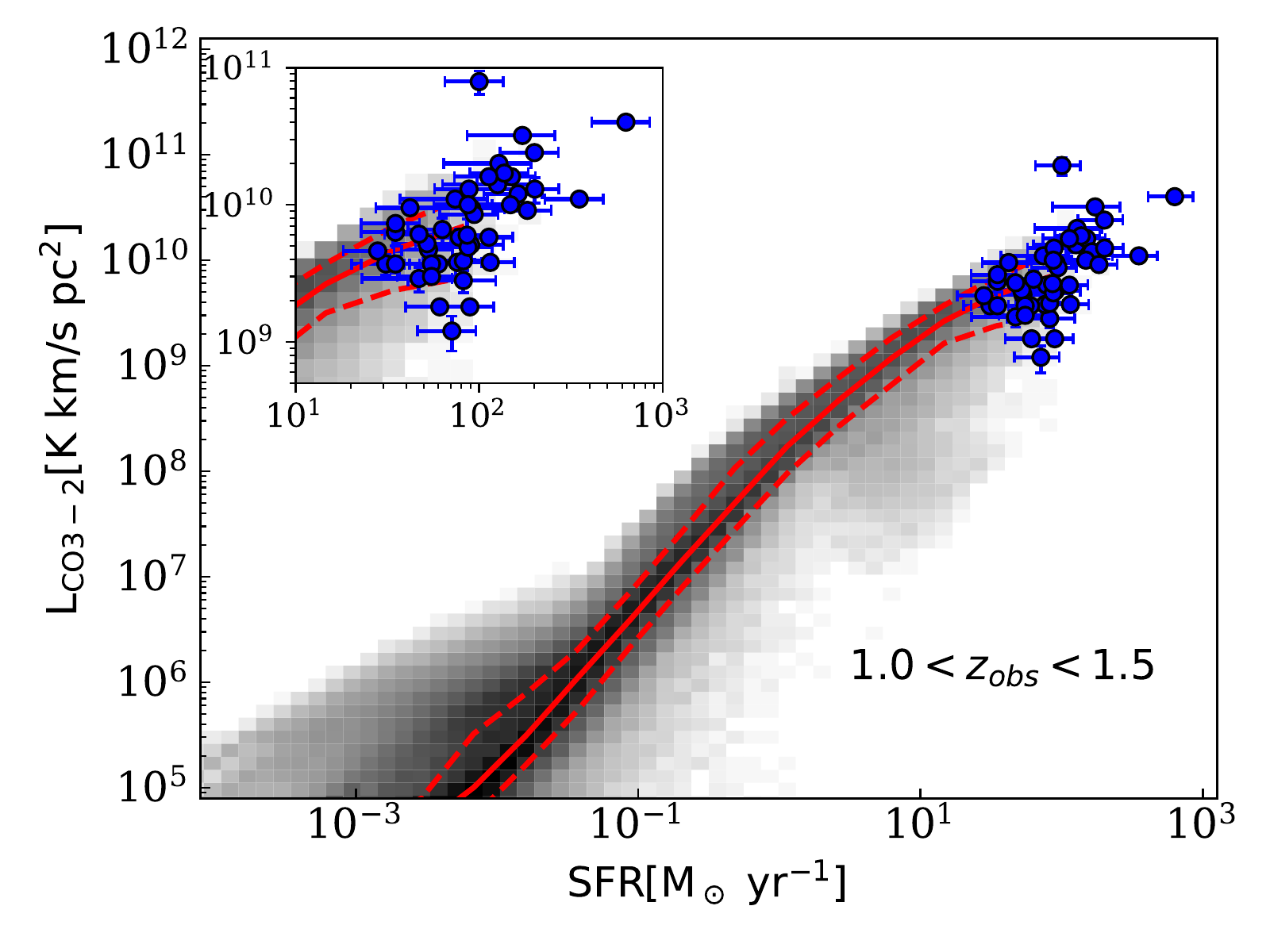}
    \xincludegraphics[clip,label=4),fontsize=\Large,width=0.49\textwidth]{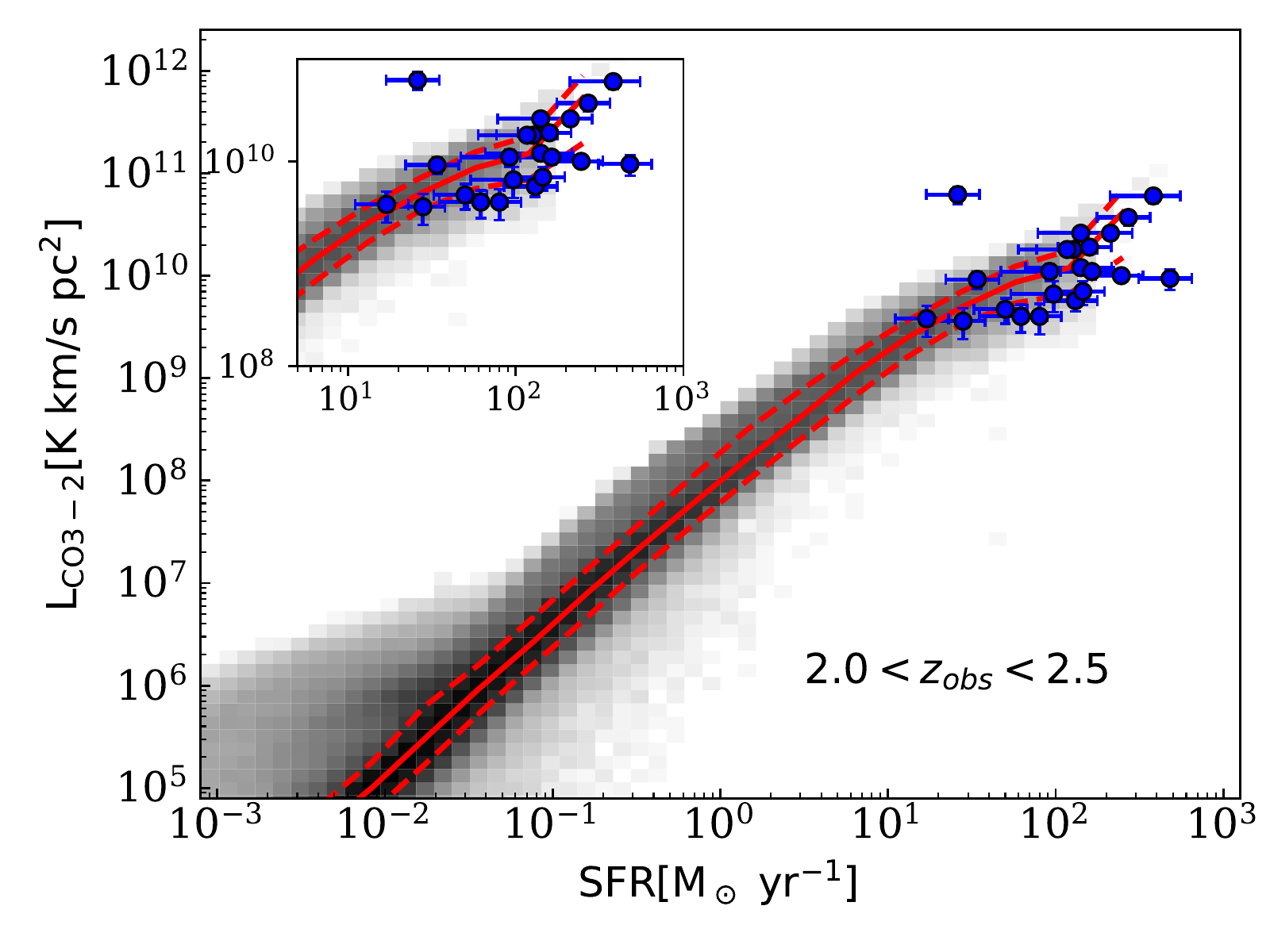}    
    \caption{Fine structure and molecular line luminosity versus galaxy SFR. The first row shows the [\cii]\ luminosity of galaxies as a function of their SFR at different redshifts. [\cii]\ observations in panel 1) are from \cite{2018MNRAS.481.1976Z}. Observations in panel 2) are provided by \cite{2015Natur.522..455C,2020arXiv200200962B}
    . The second row shows the CO J=3-2 luminosity of galaxies as a function of their SFR at different redshifts. CO J=3-2 observations in panel 3) and panel 4) are from \cite{2010Natur.463..781T} and \cite{2013ApJ...768...74T} respectively. Observational data are shown as blue dots. We only select central star forming galaxies which satisfy $\mathrm{sSFR}>1/(3t_\mathrm{H}(z))$. The upper, medium and lower red curve show the 14th, 50th and 86th percentile of line luminosity respectively predicted by the SAM+sub-mm SAM. The joint distribution of luminosity versus SFR of the mock is shown by the grey 2D histogram. Since the observations are mostly distributed among the high SFR ranges, a zoomed inset is added to each panel to better present the SAM-observation comparisons. Sub-mm line luminosity versus SFR relations predicted by the SAM $+$ sub-mm SAM are in good agreement with observations over a broad redshift range.}
    \label{fig:LSFR}
\end{figure*}

 
We present Santa Cruz SAM and sub-mm SAM predictions of [\cii]\ and CO J=3-2 luminosity versus SFR relations in different redshift ranges in Figure \ref{fig:LSFR}. In Figure \ref{fig:LSFR} we only include central galaxies of the mock lightcone that satisfy sSFR $>$ 1/(3t$_\mathrm{H}(z)$). Here sSFR is the galaxy specific SFR defined as the ratio between SFR and stellar mass, and t$_\mathrm{H}(z)$ is the Hubble time at the galaxy redshift. This selection criterion picks out star-forming galaxies which are comparable to the individual sources targeted in observed samples. We present the 14th, 50th and 86th percentile of line luminosity in each plot. [\cii]\ observational data is provided by \cite{2018MNRAS.481.1976Z} for $1.7<z_{\mathrm{obs}}<2.0$ and \cite{2015Natur.522..455C,2016MNRAS.462L...6K,2015ApJ...807..180W,2017Natur.545..457D,2014ApJ...784...99G,2013ApJ...771L..20K,2016ApJ...829L..11P,2017ApJ...836L...2B,2015A&A...576L...2S,2015MNRAS.452...54M,2014ApJ...792...34O,2016Sci...352.1559I,2017MNRAS.466..138K,2018ApJ...854L...7C} for $5.0<z_{\mathrm{obs}}<7.6$. CO J=3-2 observational data is provided by \cite{2010Natur.463..781T} for $1.0<z_{\mathrm{obs}}<1.5$ and \cite{2013ApJ...768...74T} for $2.0<z_{\mathrm{obs}}<2.5$. This comparison shows that the sub-mm SAM fiducial model reproduces [\cii] and CO J=3-2 luminosity observed in broad redshift ranges within $1<z<8$. We also check by running the SAM on a grid of halo masses, in which an equal number of realizations of halos of each mass is simulated, that the updated sub-mm SAM successfully reproduces [\cii], CO, and [\ci] line luminosities versus SFR observed in local galaxies as well as the $L_\mathrm{CI}-L_\mathrm{IR}$ measured up to $z=4$.

\begin{figure}
    \centering
    \includegraphics[trim={0 0 0 0 },clip,width=0.49\textwidth]{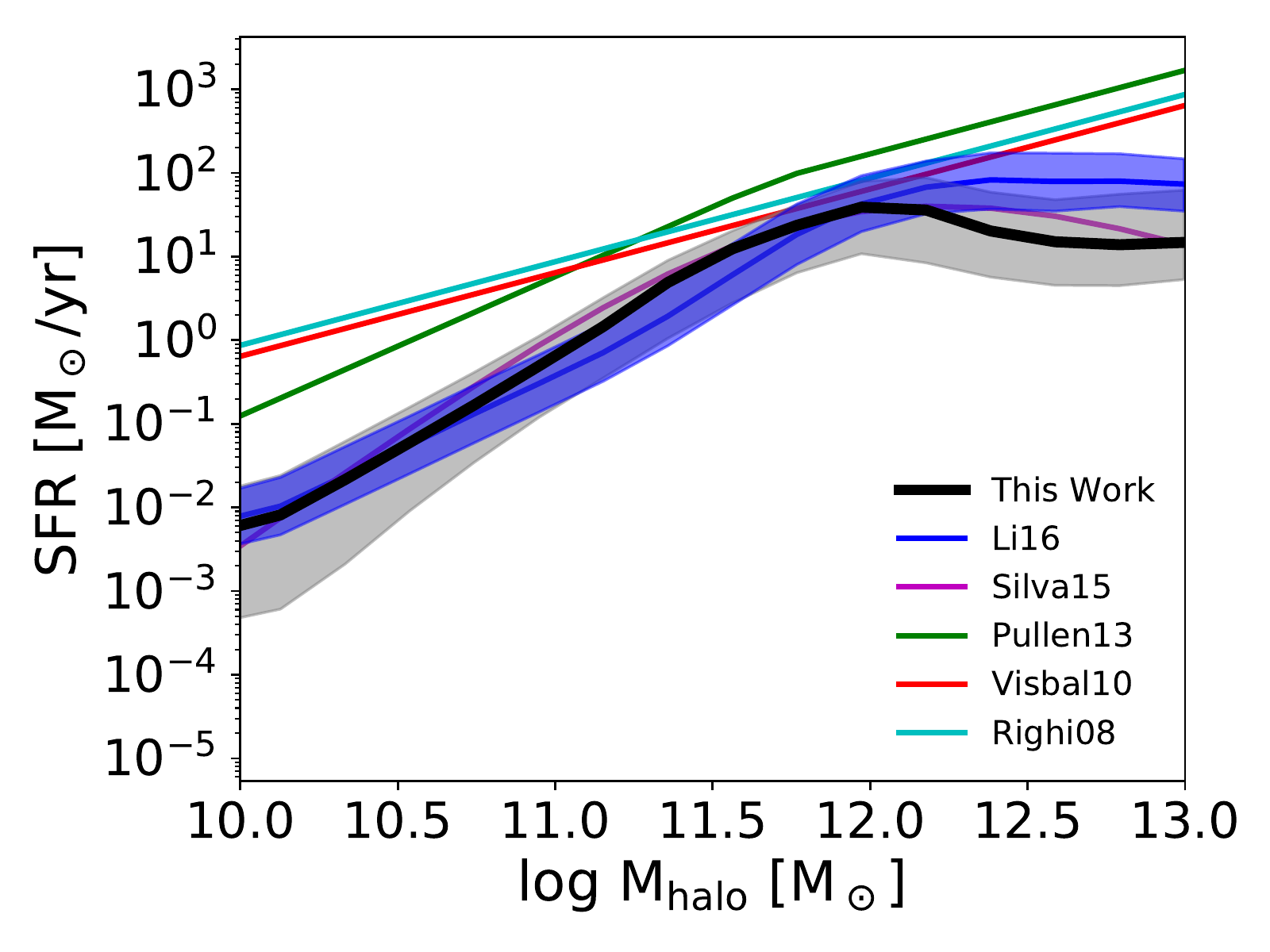}
    \includegraphics[trim={0 0 0 0},clip,width=0.49\textwidth]{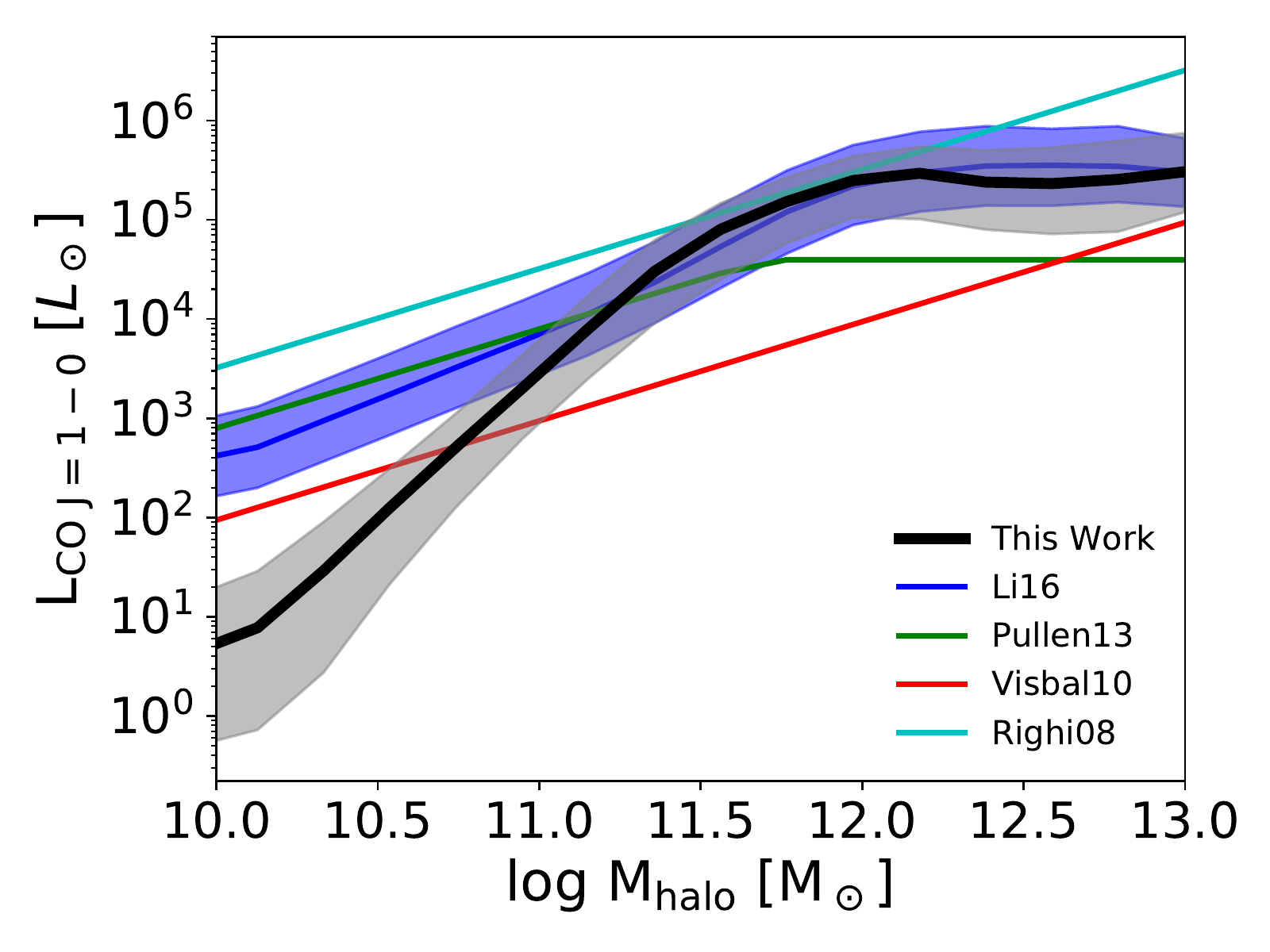}
    \caption{Top: SFR versus halo mass scaling relation at redshift $2.4<z<2.8$. Bottom: CO J=1-0 luminosity versus halo mass scaling relation at redshift $2.4<z<2.8$. Black lines with gray shading show the median and 68\% confidence level predictions of the SAM. Other colored lines show empirical relations from the literature \cite{2008A&A...489..489R,2010JCAP...11..016V,2013ApJ...768...15P,2016ApJ...817..169L,2015ApJ...806..209S}. 
    The blue band shows the 68\% confidence level of the scaling relations predicted by \citet{2016ApJ...817..169L}. The SFR versus halo mass relation predicted by the SAM is in good agreement with Li16 and Silva15, which capture the flattening of the SFR at high halo masses caused by galaxy quenching. The difference between luminosity versus halo mass relations predicted by this work and empirical models is the most significant at low halo masses due to the lack of calibration observations for faint sources. }
    \label{fig:4}
\end{figure}

We compare our CO J=1-0 predictions with models from the literature proposed by \cite{2008A&A...489..489R,2010JCAP...11..016V,2013ApJ...768...15P,2016ApJ...817..169L} in the redshift range $2.4<z_{\mathrm{obs}}<2.8$. This redshift range covers the CO J=1-0 emitters for the COMAP fiducial survey introduced in Section \ref{sec:3}. The comparisons of the SFR versus halo mass $M_{\mathrm{halo}}$ relations and the luminosity of CO J=1-0 $L_{\mathrm{CO\ J=1-0}}$ versus $M_{\mathrm{halo}}$ relations between our simulation and CO models considered in this paper are presented in Figure \ref{fig:4}. For the SFR-$M_{\mathrm{halo}}$ comparison we also consider models proposed in \cite{2015ApJ...806..209S}. \cite{2008A&A...489..489R,2010JCAP...11..016V,2013ApJ...768...15P} all assume simple linear relations between $\log\mathrm{SFR}$ and $\log M_{\mathrm{halo}}$ with slopes and other free parameters calibrated to various observations, while \cite{2015ApJ...806..209S} and \cite{2016ApJ...817..169L} model the $\mathrm{SFR}-M_{\mathrm{halo}}$ relation as double power laws or more complex functionals to capture the SFR flatness at high halo masses caused by the quiescent galaxy population. We multiply the CO J=1-0 luminosity from \cite{2010JCAP...11..016V,2013ApJ...768...15P} by a duty cycle factor $f_{\mathrm{duty}}=10^8$ yr$/t_{\mathrm{age}}(z)$ to compute the time-averaged CO J=1-0 intensities for a consistent model comparison. Here $t_{\mathrm{age}}(z)$ is the age of the universe at redshift $z$.
The $\mathrm{SFR}-M_{\mathrm{halo}}$ relation predicted by our simulation is in better agreement with the double power law behavior. Similarly, the trend of $\mathrm{L}_{\mathrm{CO\ J=1-0}}-M_{\mathrm{halo}}$ given by our simulation is closer to the most updated CO model introduced in \cite{2016ApJ...817..169L} (hereafter Li16). However, our simulation predicts lower $\mathrm{L}_{\mathrm{CO\ J=1-0}}$ at $M_{\mathrm{halo}}<10^{11}M_\odot$.\par 
\begin{figure}
    \centering
    \includegraphics[trim={0 0 0 0 },clip,width=0.49\textwidth]{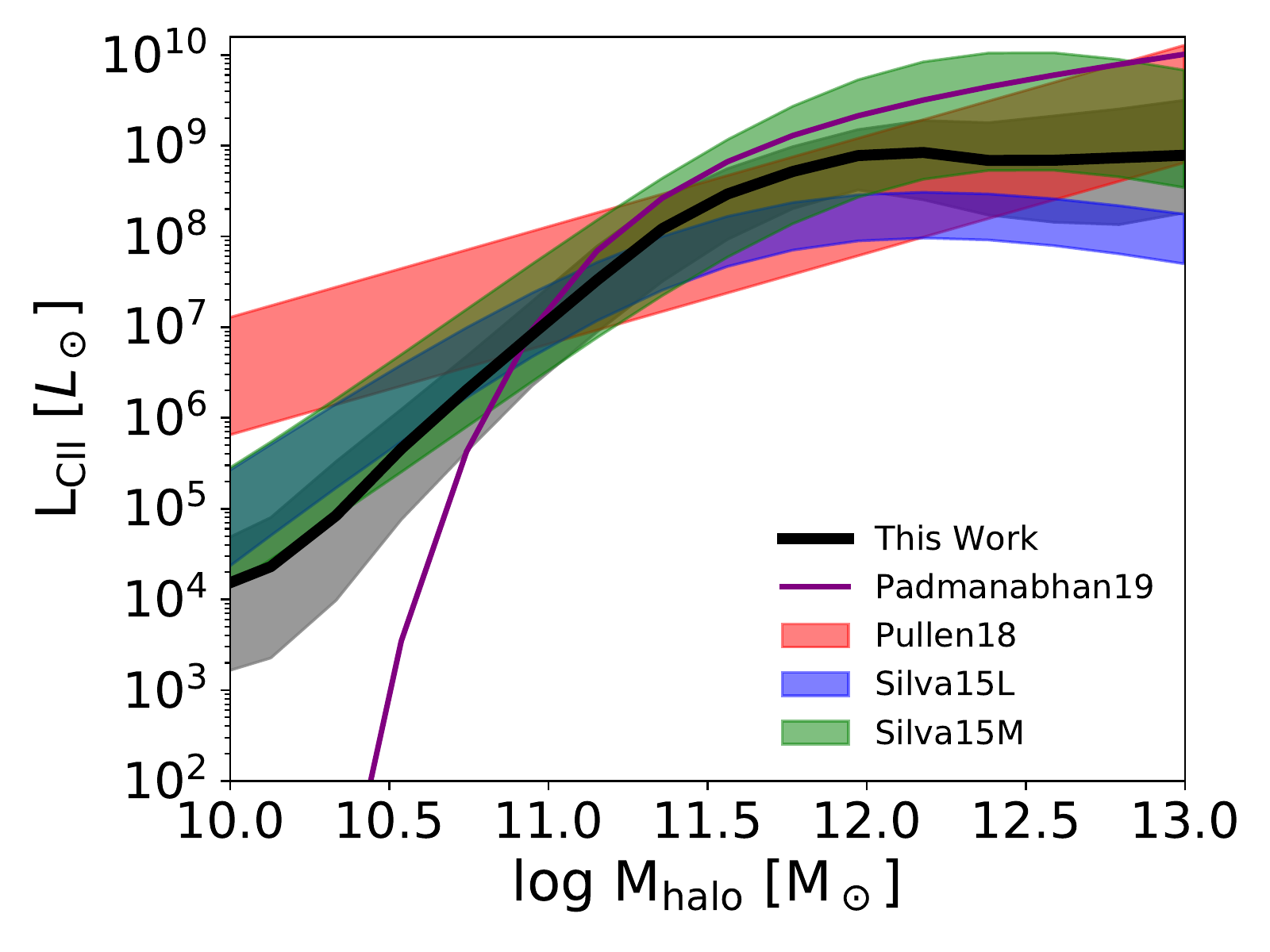}
    \caption{[\cii] luminosity versus halo mass scaling relation at redshift $2.5<z<3.5$. The empirical [\cii] relations are from \cite{2015ApJ...806..209S,2018MNRAS.478.1911P,2019MNRAS.488.3014P}. The grey band shows the 68\% confidence level of the scaling relations predicted by this work. The upper and lower bounds of bands corresponding to the Pullen18 and Silva15M models are predicted assuming ISM parameters \{$T_k^e\rightarrow\infty$, $n_e\rightarrow\infty$\} and \{$T_k^e=100$ K, $n_e=1$ cm$^{-3}$\} respectively. The upper and lower bounds of the green band correspond to the Silva15L ``$m_1$" and ``$m_4$" model respectively. The difference between the luminosity versus halo mass relations predicted by this work and empirical models in the literature are most significant for $M_\mathrm{halo}<10^{11.5}M_\odot$ and $M_\mathrm{halo}>10^{12}M_\odot$.}
    \label{fig:LCII-M}
\end{figure}

 We compare the [\cii] luminosity-$M_{\mathrm{halo}}$ relation predicted by the Santa Cruz SAM $+$ sub-mm SAM in the redshift range $2.5<z<3.5$, which covers the [\cii] emitter redshift range of the EXCLAIM fiducial survey, with the [\cii] models proposed by \cite{2015ApJ...806..209S,2018MNRAS.478.1911P,2019MNRAS.488.3014P} at $z=3$ in Figure \ref{fig:LCII-M}. \cite{2015ApJ...806..209S} introduces two [\cii] models. One (Silva15M) is based on \cite{2012ApJ...745...49G} (Gong12) but with an improved galaxy hot gas metallicity model calibrated to semi-analytic models \citep{2007MNRAS.375....2D,2011MNRAS.413..101G}, while the other (Silva15L) is a combination of empirical $L_{\mathrm{CII}}-$SFR relations calibrated to local as well as high redshift observations \citep{2001ApJ...561..766M,2013ApJ...771L..20K,2014ApJ...784...99G,2014ApJ...792...34O,2014A&A...568A..62D} and a SFR model constructed with the previously mentioned semi-analytic models \citep{2007MNRAS.375....2D,2011MNRAS.413..101G}. The [\cii] model of \cite{2018MNRAS.478.1911P} (Pullen18) is also a modified version of Gong12 but with a different [\cii] number density model. Both Silva15M and Pullen18 require ISM gas temperature $T_k^e$ and electron number density $n_e$ to determine the level abundance of [\cii] $^2\mathrm{P}_{3/2}$ and further predict [\cii] luminosity. In Figure \ref{fig:LCII-M} the upper bounds of Silva15M and Pullen18 corresponds to $T_k^e\rightarrow\infty$ and $n_e\rightarrow\infty$, while the lower bounds are predicted assuming $T_k^e=100$ K and $n_e=1$ cm$^{-3}$. The upper bound and lower bound of Silva15L correspond to the ``$m_1$" and ``$m_4$" $L_{\mathrm{CII}}$-SFR model specified in Table 1 of \cite{2015ApJ...806..209S}. \cite{2019MNRAS.488.3014P} adopt a power law form for the $L_{\mathrm{CII}}-M_{\mathrm{halo}}$ relation with an exponential cutoff, which is significantly different from the Santa Cruz SAM $+$ sub-mm SAM simulation predictions over most of the host halo mass range studied in this work.\par
 
 According to Figure \ref{fig:4} and Figure \ref{fig:LCII-M}, the scaling relations predicted by the SAM are in best agreement with models proposed by other groups at $10^{11.5}M_\odot\leq M_\mathrm{halo}\leq10^{12}M_\odot$, where high redshift galaxy observations are available. Galaxies in less massive DM halos are generally too faint to be detected, while halos more massive than this range are rare. As a result, currently the scaling relations beyond this halo mass range are not well constrained and we are unable to tell which set of model predictions is more reliable. However, unlike empirical models that are not physics-grounded, the shape of the scaling relations predicted by the SAM is determined by the underlying physical processes included in the galaxy formation model. For example, the slope of luminosity versus halo mass relation at low halo masses is largely determined by the stellar feedback strength, while the L-SFR slope at high halo masses is influenced by the galaxy quenching treatments, connected to AGN feedback in this model. Since the upcoming LIM surveys will produce constraints on various scaling relations over a broader halo mass range, the SAM will assist in interpreting these constraints in the context of a development of better understanding about various feedback mechanisms involved in galaxy formation.

\begin{figure}
    \centering
    \includegraphics[trim={0 0 0 0 },clip,width=0.49\textwidth]{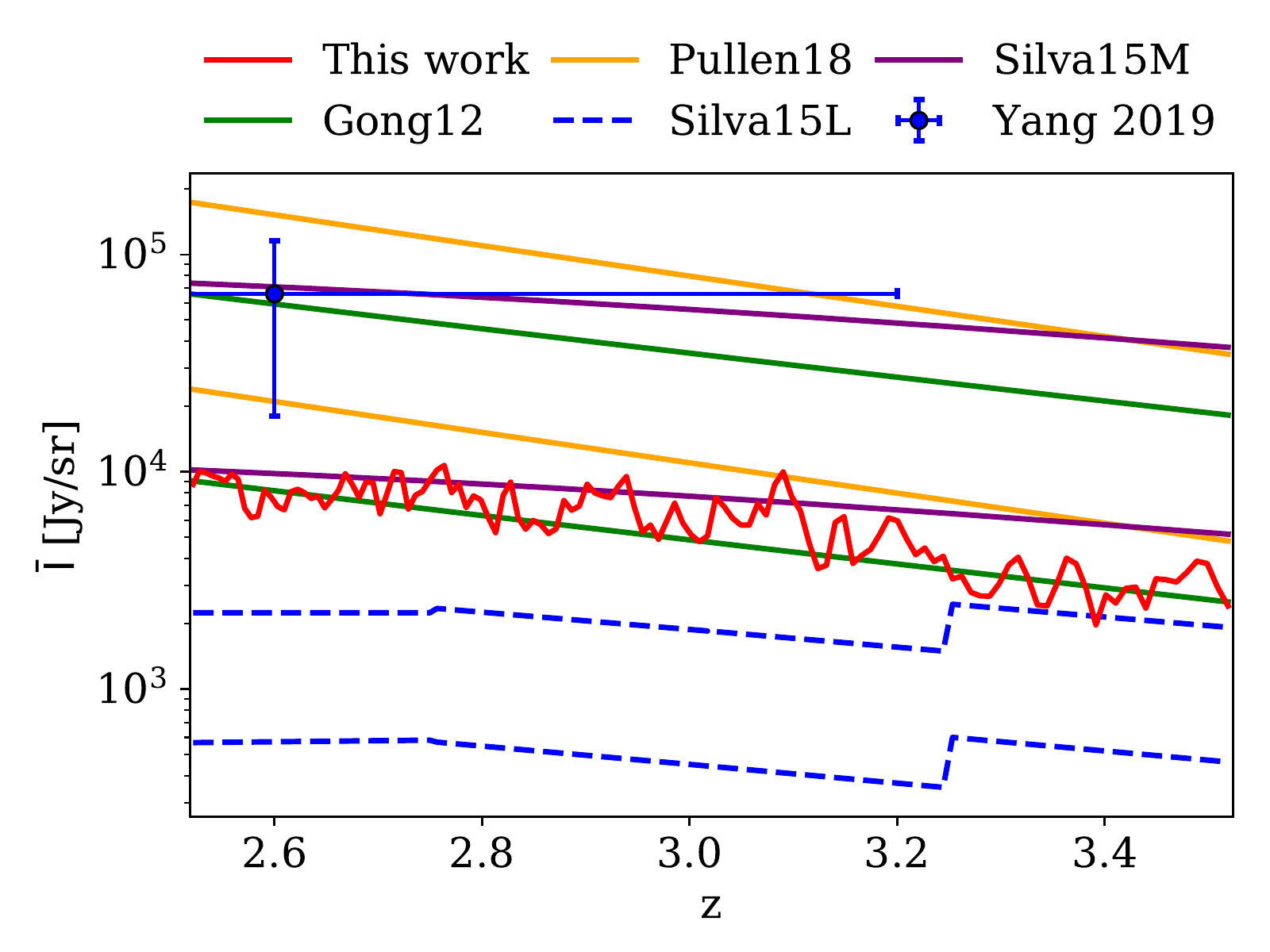}
    \caption{[\cii] intensity in the EXCLAIM fiducial survey observed frequency window predicted by the SAM, compared with models and observational constraints from \protect\cite{2012ApJ...745...49G,2015ApJ...806..209S,2018MNRAS.478.1911P,2019MNRAS.489L..53Y}. The upper and lower bounds of the Gong12, Silva15M and Pullen18 models are predicted assuming ISM parameters \{$T_k^e\rightarrow\infty$, $n_e\rightarrow\infty$\} and \{$T_k^e=100$ K, $n_e=1$ cm$^{-3}$\} respectively. The upper and lower bounds of the Silva15L models correspond to \cite{2015ApJ...806..209S} ``$m_1$" and ``$m_4$" empirical [\cii] models respectively. The SAM prediction is in best agreement with the lower bounds of collisional excitation models.}
    \label{fig:Iz}
\end{figure}

In Figure \ref{fig:Iz} we present the [\cii] intensity of the 128 EXCLAIM fiducial survey frequency channels in our EXCLAIM-like simulated map, together with predictions given by Gong12, Silva15, and Pullen18. Similar to Figure \ref{fig:LCII-M}, the upper bounds of the Gong12, Silva15M, and Pullen18 predictions assume $T_k^e\rightarrow\infty$ and $n_e\rightarrow\infty$, while the lower bounds correspond to $T_k^2=100$ K and $n_e=1$ cm$^{-3}$. The upper and lower bound of the Silva15L predictions correspond to the ``$m_1$" and ``$m_4$" model respectively. We find significant spread in predictions by different models (nearly two orders of magnitude). Among the [\cii] models presented in Figure \ref{fig:Iz}, the Silva15L empirical models assume a redshift-independent, linear $L-$SFR relation and calibrate model parameters to different galaxy samples. Models $m_1$-$m_4$ are provided to characterize the scatter in the $L$-SFR relation caused by the variation of galaxy redshift and ISM properties. In practice the $L-$SFR relation is redshift dependent. This is the main reason that the SAM [\cii] prediction at redshift $2.5<z<3.5$ is much higher than the Silva15L $m_4$ model, which is calibrated to local galaxies, while closer to $m_1$, which is calibrated to high redshift galaxy samples. Other models estimate [\cii] intensity through assuming uniform photon dominated region (PDR) environment and solving the statistical balance equation between the two [\cii] fine structure energy levels (hereafter we refer to this type of model as a collisional excitation model). Although in practice the PDR properties are complex and non-uniform, the good agreement between SAM predictions and the lower bounds of collisional excitation models indicates that on average the molecular clouds generated by the sub-mm SAM have typical gas density $n_\mathrm{H}\sim 1$ cm$^{-3}$ and temperature $T_k^e\sim100$ K. Additionally, our predicted intensity signal is more than a factor of ten lower than the measurement of \cite{2019MNRAS.489L..53Y} (Yang19).  One possible explanation for this disagreement is that the signal excess measured by Yang19 could be a combination of [\cii] emission and continuum CIB emission that is not well captured by the simple CIB model used for parameter constraints. The large fluctuations of the SAM predictions seen in Figure \ref{fig:Iz} are caused by variations in large scale structure probed by different EXCLAIM observed frequency channels.

\begin{figure*}
\centering
\includegraphics[width=0.33\textwidth]{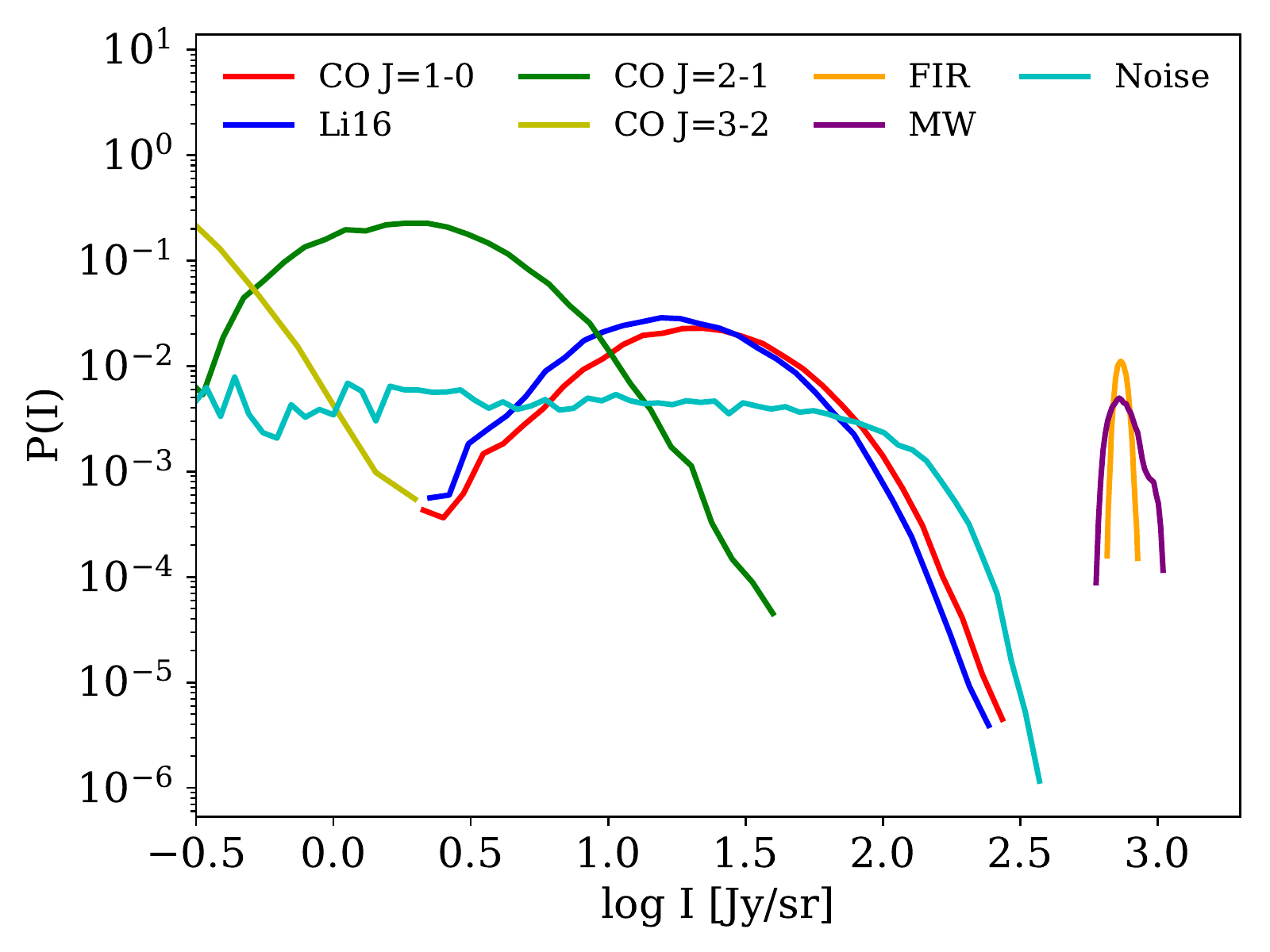}
\includegraphics[width=0.33\textwidth]{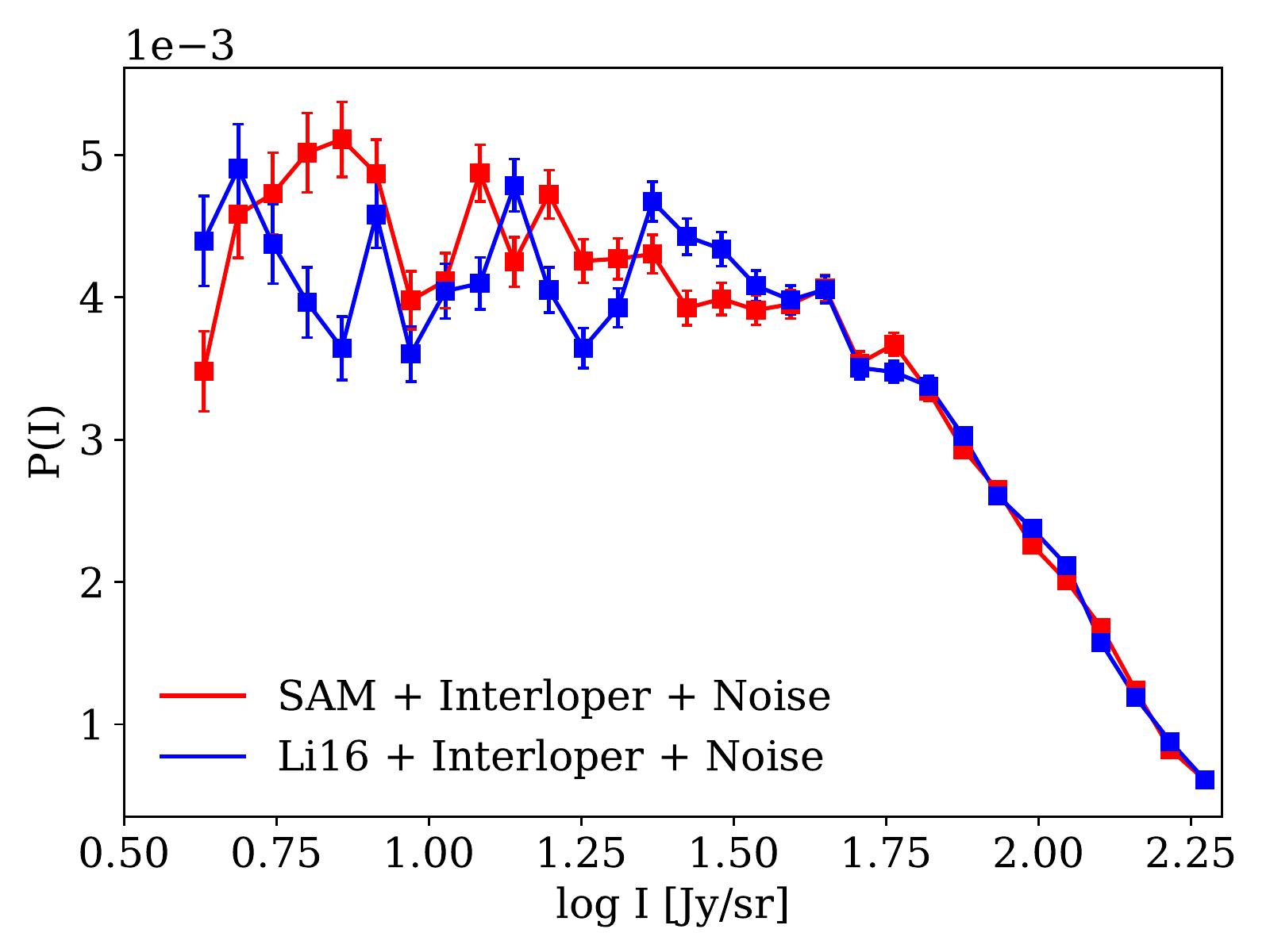}
\includegraphics[width=0.33\textwidth]{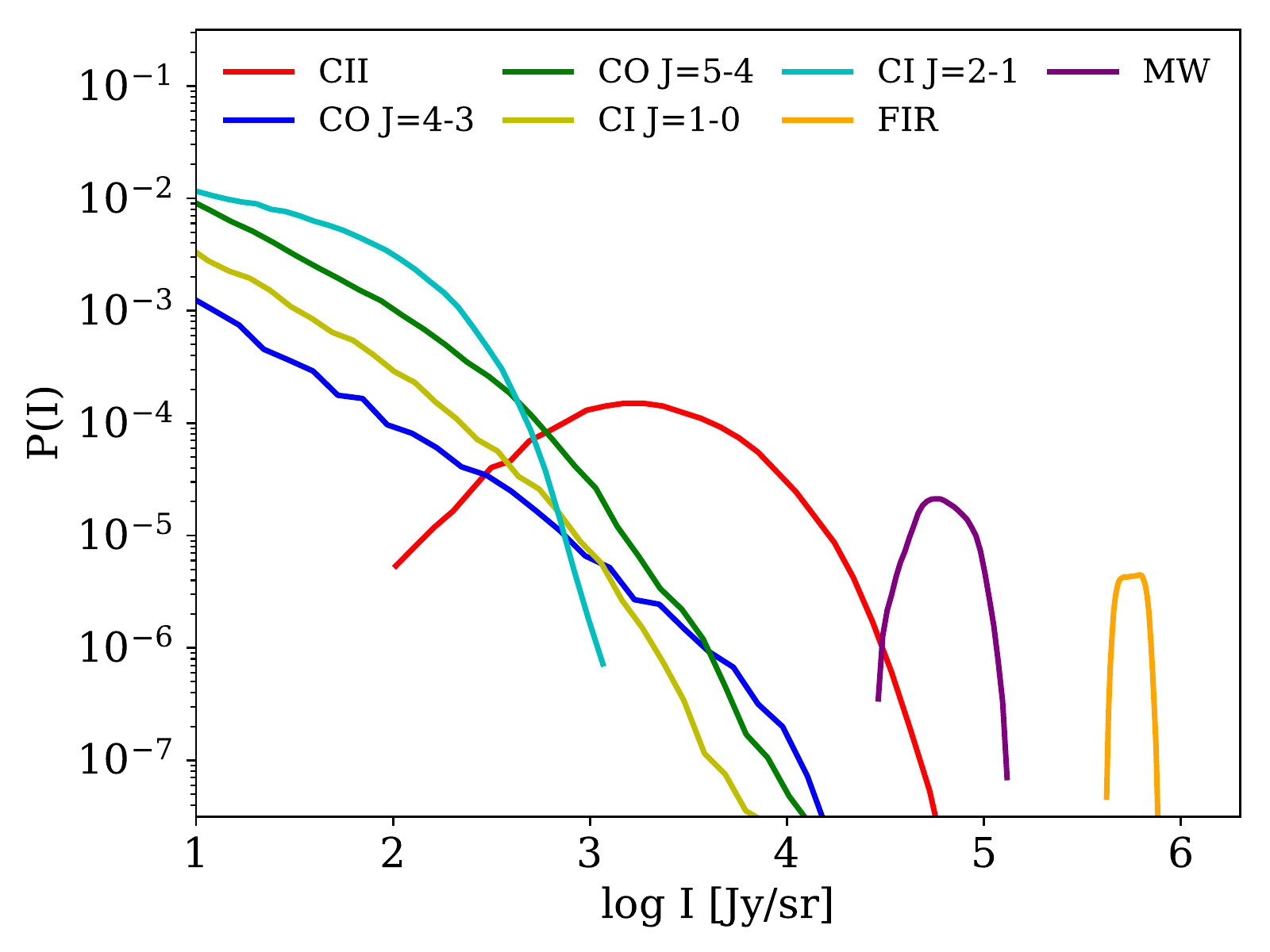}
\caption{Normalized one point PDF of the LIMs of the COMAP and EXCLAIM fiducial surveys. \textit{Left:} Line signal, continuum emission, interloper lines, and shot noise components of COMAP VID. \textit{Middle:} Comparison between COMAP VID with the CO J=1-0 signal predicted by SAM and Li16. Continuum emission is removed by subtracting the mean intensity averaged over all voxels. \textit{Right:} Line signal, continuum emission, and interloper lines of EXCLAIM VID. For the COMAP fiducial survey, the CO J=2-1 line is the primary interloper contamination source, while for the EXCLAIM fiducial survey the dominant interloper contamination comes from CO J=4-3 and CO J=5-4. The MW and dust continuum emission are much higher than the signal for both surveys. A reduced $\chi^2$ test shows that the COMAP pathfinder survey will be able to distinguish between the SAM and Li16 models with VID statistics.}
    \label{fig:1pt}
\end{figure*}

We present predictions of the normalized voxel intensity distribution (VID) or one-point PDF for our COMAP-like and EXCLAIM-like fiducial mock maps in the left and right panel of Figure \ref{fig:1pt}. Before computing the VID we downgrid the map with resolution $\theta_\mathrm{pix}$ to the LIM survey spatial resolution $\theta_\mathrm{FWHM}$ in order to minimize correlations between histogram bins \citep{2014MNRAS.440.2791V}.  This is done to ensure reliable VID noise estimation, used in the later reduced $\chi^2$ test. In the left panel of Figure \ref{fig:1pt}, we show separately the contribution to the normalized VID from CO J=1-0, as well as from the primary interlopers CO J=2-1 and CO J=3-2, for the COMAP-like map. We also show the shot noise, MW and dust continuum (CIB) components. We show a comparison with the predictions of the Li16 model, which only includes the primary CO J=1-0 signal. As we showed in Figure~\ref{fig:4},  the Li16 model input $L_{\mathrm{CO\ J=1-0}}$-$M_{\mathrm{halo}}$ relation is actually quite similar to the one that emerges from our simulation, but it has a shallower slope and higher amplitude at halo masses below $\sim 10^{11.5}$ M$_\odot$. Due to the low luminosity of less massive emitters, the higher amplitude of the $L_{\mathrm{CO\ J=1-0}}$-$M_{\mathrm{halo}}$ relation given by Li16 does not lead to an overall higher signal.
As a result, the CO J=1-0 VID of the SAM is higher than that of Li16 for $I\gtrsim30$ Jy/sr, mainly because the SAM CO J=1-0 is slightly brighter than the Li16 predictions in the halo mass range $10^{11.2}M_\odot\lesssim M_\mathrm{halo}\lesssim10^{12}M_\odot$. The CO J=2-1 line is the primary interloper line contaminant but is still about 1 order less luminous than the CO J=1-0 signal. The MW and dust continuum emission are about two orders of magnitude higher than the signal.\par
We perform a reduced $\chi^2$ test to estimate if the sensitivity of the upcoming COMAP ``pathfinder" survey is high enough to distinguish different CO J=1-0 models. We clean the bright continuum foreground from the mock intensity map by subtracting the mean intensity averaged over all voxels \citep{2017MNRAS.467.2996B}. This simple foreground removal method is not optimal since it only estimates the continuum amplitude at the first order, and it also shifts the VID by a constant after cleaning. More accurate VID continuum treatment requires further work. However, \cite{2017MNRAS.467.2996B} showed that this method can effectively clean the continuum. It does not significantly bias or weaken the VID constraints on the Schechter CO luminosity function parameters. Assuming that the intensities of all voxels are almost independent and follow a binomial distribution, the VID variance can be estimated as $\sigma^2(I)=B(I)(1-B(I)/N_\mathrm{vox})\approx B(I)$, where $B(I)$ is the number of voxels with intensity falls in the bin centered at $I$, and $N_\mathrm{vox}$ is the total voxel number of the mock LIM data.  These assumptions break down at high signal-to-noise ratios, but should suffice for the case considered here \citep{2019ApJ...871...75I}. We bin the cleaned intensity maps with CO J=1-0 signals predicted by SAM and Li16 into $N_b=30$ logarithmically spaced bins within $4\leq I/[\mathrm{Jy/sr}]\leq200$. Voxels with intensities beyond this range are ignored. The normalized VID predicted by SAM and Li16 after the continuum cleaning are presented in the middle panel of Figure \ref{fig:1pt}. The reduced $\chi^2$ between the SAM and Li16 VID is:
\begin{equation}
    \chi^2_\nu=\dfrac{1}{N_b}\sum\dfrac{(B^\mathrm{Li16}(I)-B^\mathrm{SAM}(I))^2}{\sigma_\mathrm{SAM}^2(I)/4}=6.8\,,
\end{equation}
where $B(I)$ consists CO J$=1-0$, interloper lines and instrumental noise. We reduce the VID variance by a factor of 4 because the COMAP survey measures 4 sky patches. This shows that COMAP pathfinder fiducial survey can distinguish SAM $+$ sub-mm SAM and Li16 models with the LIM VID statistics.\par
 
For the EXCLAIM survey (shown Figure \ref{fig:1pt} right panel), the CIB is the brightest continuum background. CO J=4-3 and CO J=5-4 are the most significant interloper lines.\par 


\begin{figure*}
\centering
\includegraphics[width=0.33\textwidth]{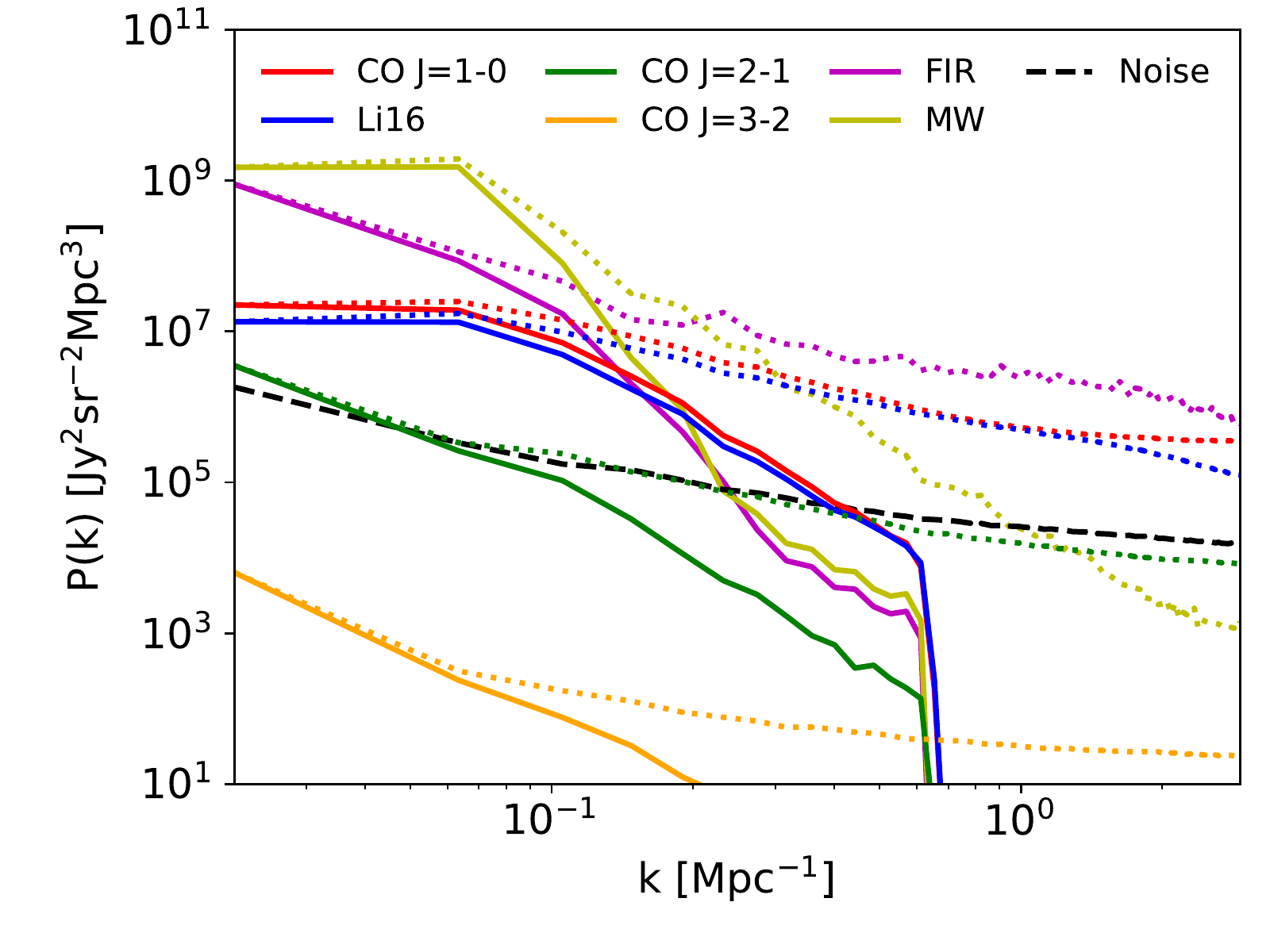}
\includegraphics[width=0.33\textwidth]{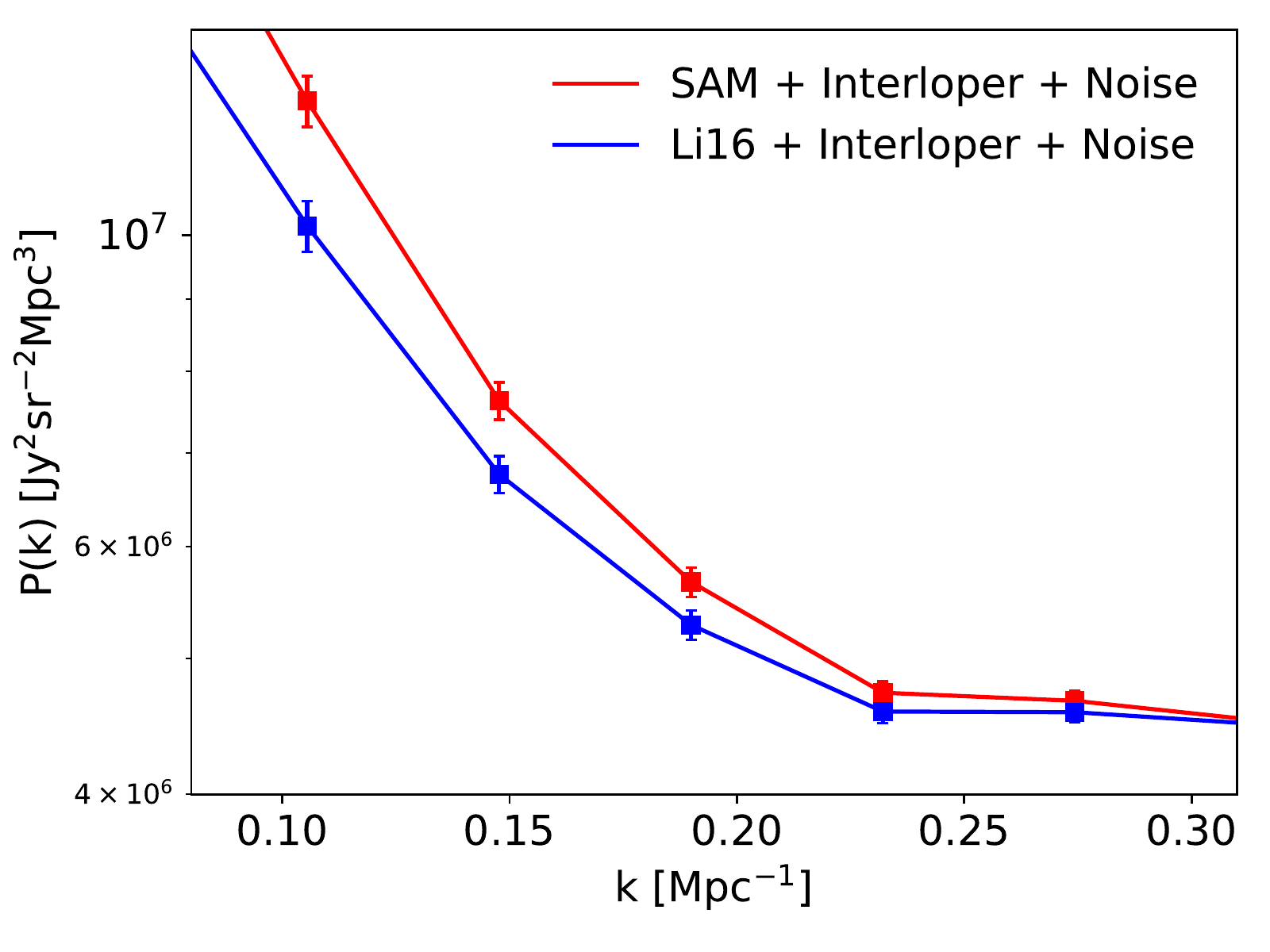}
\includegraphics[width=0.33\textwidth]{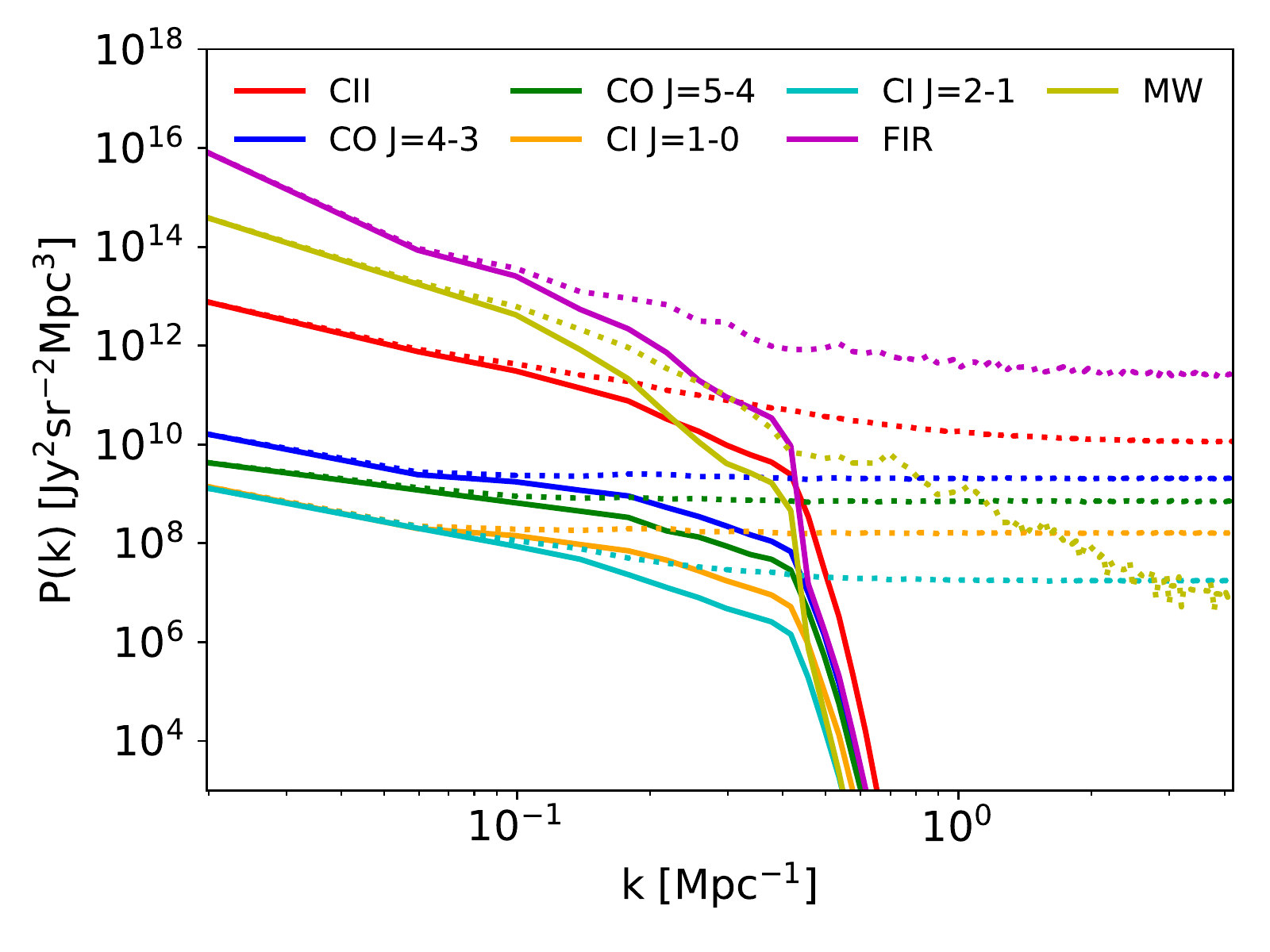}
\caption{Fiducial COMAP power spectrum and EXCLAIM power spectrum. Solid (Dotted) curves show the emission line power spectra with (without) the map smoothing representing the beam. The black dashed curve shows the $1\sigma$ error contributed by the COMAP instrumental noise. \textit{Left}: Line signal, continuum emission, interloper lines, and shot noise components of COMAP power spectrum. \textit{Middle:} Comparison between COMAP power spectrum with the CO J=1-0 signal predicted by SAM and Li16. The continuum is estimated and subtracted by fitting the continuum intensity dependence of frequency by a second order log-log polynomial. \textit{Right:} Line signal, continuum emission, and interloper lines of EXCLAIM power spectrum. We confirm that with the presence of interloper and instrumental noise contamination, the COMAP pathfinder fiducial survey is able to distinguish SAM $+$ sub-mm SAM and the empirical model Li16 through the auto CO J=1-0 power spectrum.}
    \label{fig:Pk}
\end{figure*}

Finally, we show the spherically averaged power spectrum of the fiducial COMAP and EXCLAIM mock surveys in the left and right panel of Figure \ref{fig:Pk}. Dotted lines show the power spectrum of emission lines for the mock LIM with angular resolution one order of magnitude smaller than the beam width of the corresponding LIM survey, while solid curves account for the power spectrum attenuation caused by smoothing. The black dashed line shows the $1\sigma$ power spectrum error contributed by the COMAP instrumental noise:
\begin{equation}
    \sigma_n(k)=\dfrac{P_n(k)}{\sqrt{N_\mathrm{modes}}}\,,
\end{equation}
Here $P_n$ is the power spectrum of the instrumental noise, and $N_\mathrm{modes}$ is the number of $k$ modes in each $k$ bins. We reduce the instrumental noise for the COMAP pathfinder fiducial survey by a factor of 2 since the power spectrum will be averaged over four independent sky patches in practice.\par 
Similar to the VID foreground cleaning process, the strong continuum emissions presented in the power spectrum can be removed through taking advantage of their smooth frequency spectra. In this work we follow \cite{2006ApJ...650..529W} and estimate the continuum intensity dependence of frequency by a second order log-log polynomial for the foreground removal. We confirm that the continuum components in the power spectrum are effectively removed at $k>0.1$Mpc$^{-1}$. In the middle panel of Figure \ref{fig:Pk} we compare the cleaned power spectrum with the CO J=1-0 signal simulated by SAM and Li16. We estimate the variance of the power spectrum as $\sigma^2(k)=P^2(k)/N_\mathrm{mode}$, where $P(k)$ is the spherically averaged power spectrum consisting smoothed signal, smoothed interloper emission and instrumental noise, and compute the reduced $\chi^2$ between the Li16 model prediction and SAM $+$ sub-mm SAM over the $N_b=5$ $k$ bins within $0.1 \leq k/[\mathrm{Mpc^{-1}}]\leq 0.3$:
\begin{equation}
    \chi^2_\nu=\dfrac{1}{N_b}\sum\dfrac{(P^\mathrm{Li16}(k)-P^\mathrm{SAM}(k))^2}{\sigma^2_\mathrm{SAM}/4}=9.1\,.
\end{equation}
Here again we reduce the power spectrum variance by a factor of 4 due to the 4 independent sky patches measured by COMAP survey. We ignore the power spectrum at $k>0.3$Mpc$^{-1}$ since the shot noise dominates at small scale. The $\chi^2_\nu$ test result shows that the COMAP pathfinder fiducial survey can distinguish the Li16 model and the SAM $+$ sub-mm SAM. Ultimately, we want to use the LIM summary statistics to constrain physical processes in galaxy formation. The ability of the LIM statistics to discriminate between two models that both fit observations of bright sub-mm line emission sources, but differ in the mapping between DM halo mass and line luminosity for lower mass halos, demonstrates the promise of this approach.

\section{Conclusion}\label{sec:4}
In this work we present a framework for constructing synthetic multi-tracer LIM maps, based on a mock lightcone extracted from an N-Body simulation and populated with a physics-based semi-analytic galaxy formation model. The workflow is: 1) Construct DM halo lightcone from N-body simulation catalog. 2) Use Santa Cruz SAM to simulate the DM halo merger history and generate galaxy formation histories. 3) Use sub-mm SAM to estimate the luminosity of [\cii], CO, [\ci]\ lines for each simulated galaxy. 4) Grid the discrete galaxy and emission line catalog in the [RA, DEC, $\nu_{\mathrm{obs}}$] space and generate 3D intensity maps. Following this procedure, we have constructed a mock lightcone which covers a 2 deg$^2$ sky area and extends over the redshift range $0\leq z\leq10$. We check that the emission line luminosity versus SFR relations for [\cii] and CO J=3-2 predicted by the mock lightcone are in good agreement with observations at various cosmic times. The integrated CIB spectrum predicted by the SAM is also consistent with observational constraints over a wide frequency range.\par 
We show that the widely used scaling relations such as SFR-M$_{\mathrm{halo}}$ and L-M$_{\mathrm{halo}}$ predicted by our simulation are in good agreement with empirical models in the  literature in the halo mass range $10^{11.5}M_\odot\leq M_\mathrm{halo}\leq10^{12} M_\odot$. However, the differences become significant beyond this halo mass range, where no observations are currently available and the scaling relations are not well constrained. Due to the physics-based nature of the SAM $+$ sub-mm SAM approach, future LIM observations will constrain scaling relations over wider halo mass ranges and provide more information about important mechanisms in the galaxy formation process, such as stellar and AGN feedback.\par 
Based on this mock lightcone, we simulate intensity maps for two fiducial LIM experiments with instrumental parameters aligned with the upcoming COMAP and EXCLAIM LIM surveys. Our simulation shows that the MW and CIB continuum emission are 2-3 orders of magnitude brighter than the signal. CO lines are the dominant interloper contamination sources for both the COMAP and EXCLAIM surveys. We also show that with the presence of instrumental noise and interloper contamination, the CO J=1-0 line auto power spectrum predicted by SAM $+$ sub-mm SAM is significantly different from the prediction of the Li16 empirical model. 
The ability of LIM summary statistics to discriminate between models that are calibrated/validated on bright sources but differ in the predicted properties of fainter objects demonstrates the promise of this approach for constraining physical processes in galaxy evolution. 
There are well studied frameworks on using one point LIM PDF, conditional voxel intensity distribution or power spectrum to constrain high redshift galaxy properties such as the line luminosity function and star formation rate density. We will use the 2 deg$^2$ mock lightcone and intensity maps simulated in this work to test different methods to extract these physical quantities, as well as foreground removal techniques, in future works. Although we only constructed mock LIM data for the COMAP and EXCLAIM surveys as two examples in this work, this simulation framework can be easily applied to other upcoming LIM surveys.

\section{Acknowledgement}
We thank Trevor M. Oxholm for beneficial discussions about the EXCLAIM instrumental noise estimation. S. Y thanks L. Y. Aaron Yung for SAM usage guidance. We gratefully acknowledge support from the Simons Foundation. This work made use of the Flatiron Institute Computing Cluster. ARP was supported by NASA under award numbers 80NSSC18K1014 and NNH17ZDA001N. 

\appendix
\section{Minimum halo mass}\label{appendix:Mmin}
Here we show that halos with mass less than $10^{10}M_\odot$ make a negligible contribution to the average line intensity (and similarly, to the other LIM statistics) for all ISM sub-mm emission models with a steep low mass slope, including the sub-mm SAM used in this work.\par
\begin{figure}
    \centering
    \includegraphics[width=0.45\textwidth]{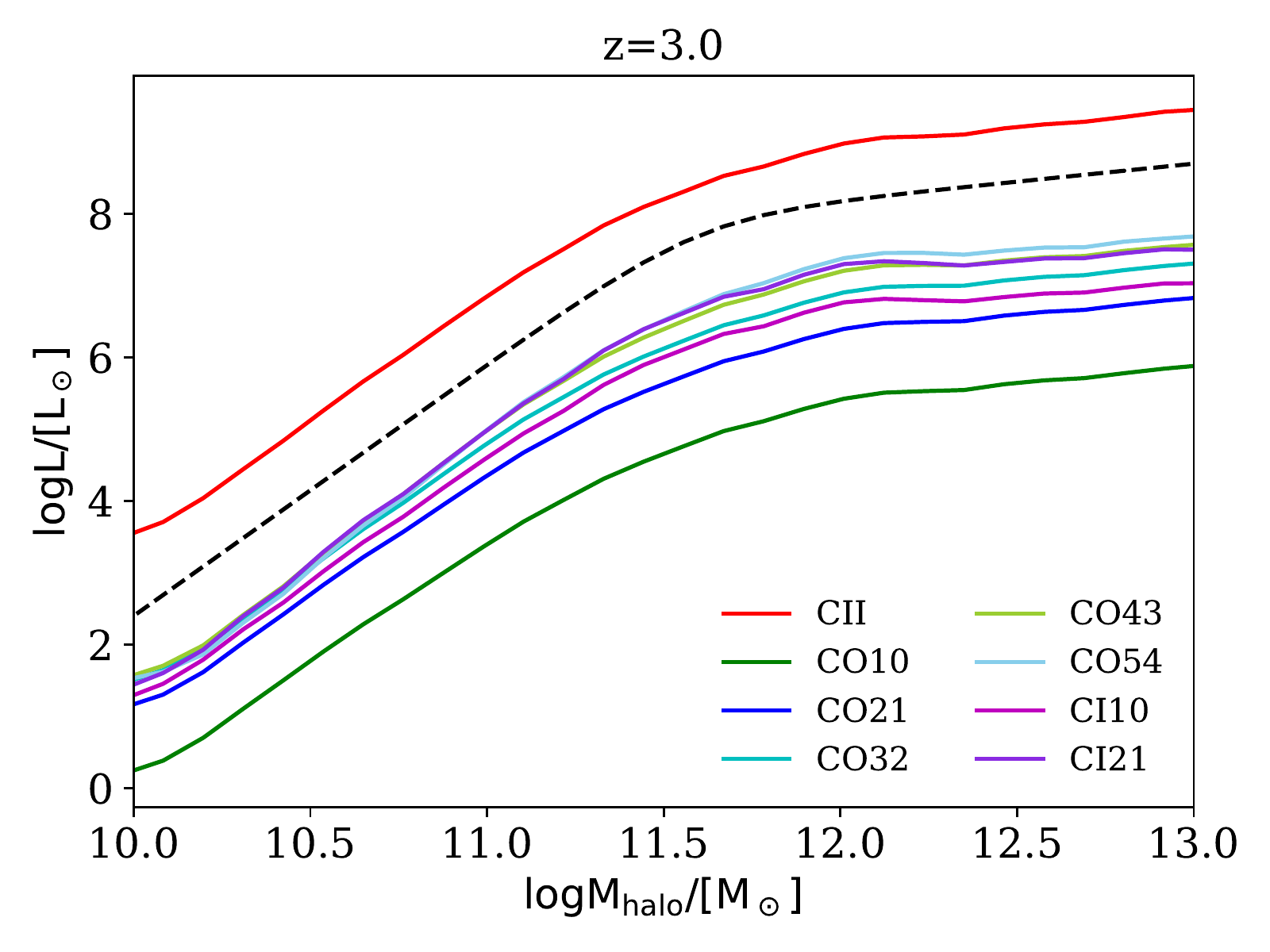}
    \includegraphics[width=0.45\textwidth]{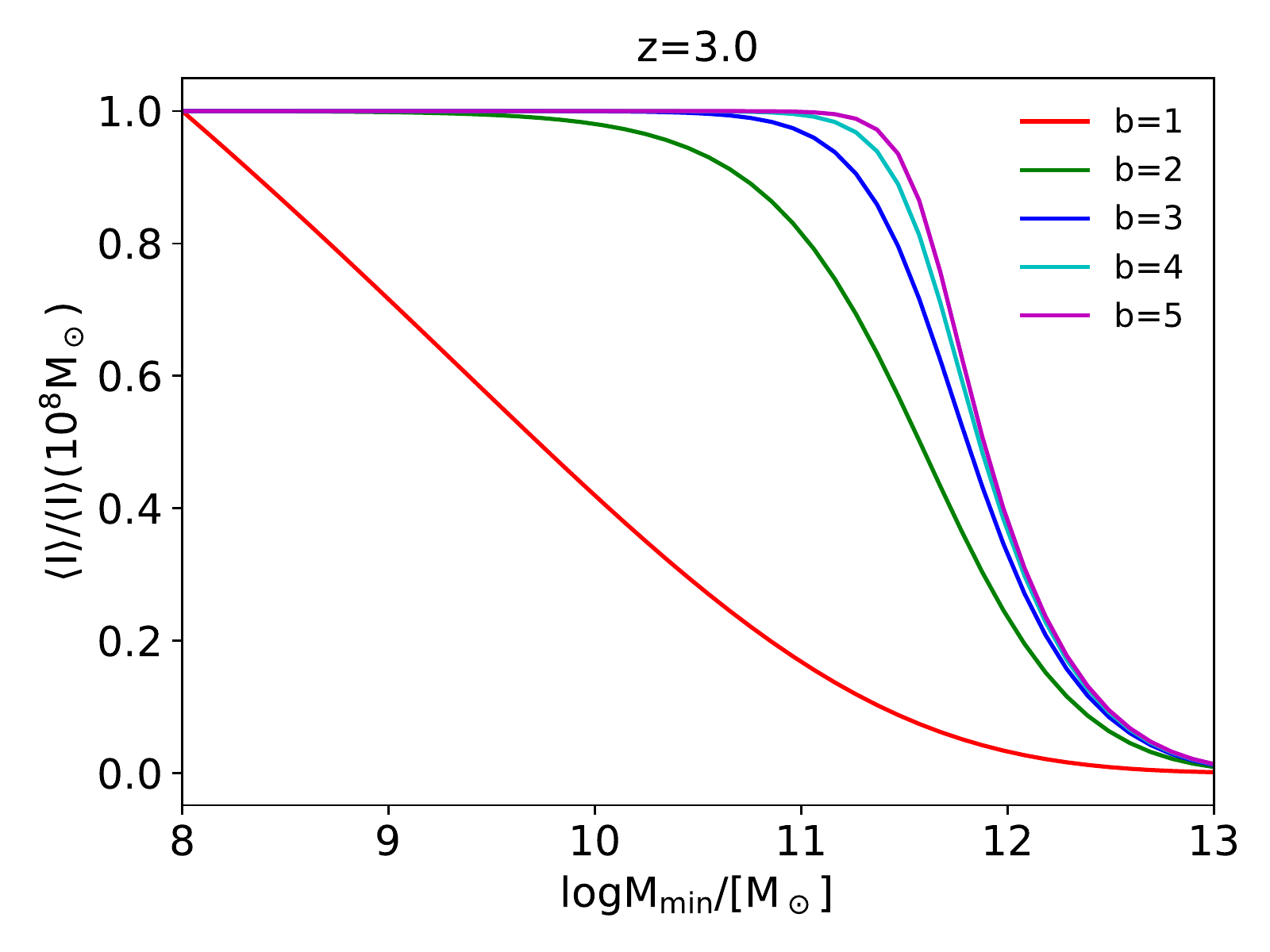}
    \caption{Left: Line luminosity versus halo mass relations $L(M)$. Mean $L(M)$ relations simulated by SAM are shown in solid curves. The black dashed line shows a double power law $L(M)$ relation defined in Eq (\ref{eq:doublepowerlaw}), with parameters $N=10^8L_\odot$, $M_1=10^{11.6}M_\odot$, $a=-0.5$, and $b=3.5$. Right: Fractional loss of averaged line intensity as a function of the lower bound of halo mass integration. Low halo mass slope $b$ of $L(M)$ relation is varied. Other double power law parameters are fixed as $N=10^8L_\odot$, $M_1=10^{11.6}M_\odot$, and $a=-0.5$.}
    \label{fig:LM}
\end{figure}

Most of the current line luminosity versus halo mass relations, including the SAM simulation results, can be approximated by double power law relations:
\begin{equation}\label{eq:doublepowerlaw}
    L(M)=N\left[\left(\dfrac{M}{M_1}\right)^{-b}+\left(\dfrac{M}{M_1}\right)^a\right]^{-1}\,.
\end{equation}
In Figure \ref{fig:LM} left panel we present the mean SAM $L(M)$ relations at redshift $z=3$, together with a double power law relation $L(M)/[L_\odot]=10^8[(M/[M_\odot]/10^{11.6})^{-3.5}+(M/[M_\odot]/10^{11.6})^{-0.5}]^{-1}$. Here we ignore the scatter in the $L(M)$ relation because it will not influence the average emission line intensity. Further study shows that the $a$ and $b$ parameters, which characterize the SAM $L(M)$ slopes for high and low halo mass ranges, do not vary significantly with redshift.\par
For cases where the low mass halos in the system are negligible, the assumed $L(M)$ models have a relatively large value for the $b$ parameter, in which case the halo luminosity decreases very rapidly as the halo mass decreases. In Figure \ref{fig:LM} right panel we show the fractional loss of mean intensity as a function of the halo mass integration lower bound:
\begin{equation}
    \dfrac{\langle I\rangle(M_\mathrm{min})}{\langle I\rangle(10^8M_\odot)}=\dfrac{\int_\mathrm{M_{min}}^\infty dM(dn/dM)L(M)}{\int_{10^8M_\odot}^\infty dM(dn/dM)L(M)}\,.
\end{equation}
Here the $L(M)$ models are the double power law defined in Eq (\ref{eq:doublepowerlaw}) with fixed parameters $N=10^8\ [L_\odot]$, $M_1=10^{11.6}\ [M_\odot]$ and $a=-0.5$. We adopt the halo mass function $dn/dM$ given by the model of \cite{2002MNRAS.329...61S}. This toy model test shows that when the $L(M)$ relation low mass slope $b>2$, halos with mass less than $10^{10}$ $M_\odot$ make negligible contribution to the average intensity. Notice that $L(M)$ models such as \cite{2010JCAP...11..016V,2011ApJ...741...70L} assume linear relations. i.e. $b=1$. For those linear $L(M)$ models the low mass halos contribute significantly to the average line intensity. However, SAM predictions prefer $b\approx 3.5$ for all the emission lines probed in this paper, and we therefore confirm that setting mass resolution of our simulations as $10^{10}M_\odot$ will not significantly influence the LIM statistics.  
\par


\bibliography{sample63}{}
\bibliographystyle{aasjournal}



\end{document}